\documentclass[11pt,titlepage]{article} 
\usepackage{amsfonts}
\usepackage{amsmath}
\usepackage{amssymb}
\usepackage{tipa}
\usepackage{url}
\usepackage[super,sort&compress]{natbib}
\usepackage{caption}
\usepackage{graphicx,rotating}
\usepackage [ table ]{ xcolor }
\usepackage{setspace}
\usepackage[table]{xcolor}
\begin{document}
\title{Joint analysis of functional genomic data and genome-wide association studies of 18 human traits}
\author{Joseph K. Pickrell$^{1,2}$\\
\small $^1$ New York Genome Center, New York, NY\\
\small $^2$ Department of Biological Sciences, Columbia University, New York, NY \\
\small Correspondence to: \url{jkpickrell@nygenome.org}
}
\maketitle
\begin{abstract}
Annotations of gene structures and regulatory elements can inform genome-wide association studies (GWAS). However, choosing the relevant annotations for interpreting an association study of a given trait remains challenging. We describe a statistical model that uses association statistics computed across the genome to identify classes of genomic element that are enriched or depleted for loci that influence a trait.  The model naturally incorporates multiple types of annotations. We applied the model to GWAS of 18 human traits, including red blood cell traits, platelet traits, glucose levels, lipid levels, height, BMI, and Crohn's disease. For each trait, we evaluated the relevance of 450 different genomic annotations, including protein-coding genes, enhancers, and DNase-I hypersensitive sites in over a hundred tissues and cell lines. We show that the fraction of phenotype-associated SNPs that influence protein sequence ranges from around 2\% (for platelet volume) up to around 20\% (for LDL cholesterol); that repressed chromatin is significantly depleted for SNPs associated with several traits; and that cell type-specific DNase-I hypersensitive sites are enriched for SNPs associated with several traits (for example, the spleen in platelet volume). Finally, by re-weighting each GWAS using information from functional genomics, we increase the number of loci with high-confidence associations by around 5\%. 
\end{abstract}
%\tableofcontents
\pagebreak

\section{Introduction}
A fundamental goal of human genetics is to create a catalogue of the genetic polymorphisms that cause phenotypic variation in our species and to characterize the precise molecular mechanisms by which these polymorphisms exert their effects. An important tool in the modern human genetics toolkit is the genome-wide association study, in which hundreds of thousands or millions of single nucleotide polymorphisms (SNPs) are genotyped in large cohorts of individuals and each polymorphism tested for a statistical association with some trait of interest. In recent years, GWAS have identified thousands of genomic regions that show reproducible statistical associations with a wide array of phenotypes and diseases \citep{Visscher:2012fk}. 

In general, the loci identified in GWAS of multifactorial traits have small effect sizes and are located outside of protein-coding exons \cite{Hindorff:2009uq}. This latter fact has generated considerable interest in annotating other types of genomic elements apart from exons. For example, the ENCODE project has generated detailed maps of histone modifications and transcription factor binding in six human cell lines, partially motivated by the goal of interpreting GWAS signals that may act via a mechanism of gene regulation \citep{ENCODE-Project-Consortium:2012fk}. Methods for combining potentially rich sources of functional genomic data with GWAS could in principle lead to important biological insights. The development of such a method is the aim of this paper. 

There are two lines of research that motivate our work on this problem. The first is what are often called ``enrichment" analyses. In this type of analysis, the most strongly associated SNPs in a GWAS are examined to see if they fall disproportionately in specific types of genomic region. These studies have found, for example, that SNPs identified in GWAS are enriched in protein-coding exons, promoters, and untranslated regions (UTRs) \citep{Hindorff:2009uq, Schork:2013kx} and among those that influence gene expression \citep{Nicolae:2010ys, Lappalainen:2013fk}. Further, in some cases, SNPs associated with a trait are enriched in gene regulatory regions in specific cell types \citep{Cowper-Sal-lari:2012zr, Gerasimova:2013ly, Trynka:2013ve, Ernst:2011fk, Maurano:2012vn, karczewski2013systematic, Paul:2011fk,Paul:2013uq, van2012seventy, Parker:2013fk, Global-Lipids-Genetics-Consortium:2013uq, Hnisz:2013fk} or near genes expressed in specific cell types \citep{Lui:2012qf, hu2011integrating}. However, the methods in these studies are generally not able to consider more than a single annotation at a time (with a few exceptions \cite{Kindt:2013aa, gagliano2013bayesian}). Further, they are not set up to answer a question that we find important: consider two independent SNPs with equivalent P-values of $1\times 10^{-7}$ in a GWAS for some trait (note that this P-value does not reach the standard threshold of $5\times10^{-8}$ for ``significance"), the first of which is a nonsynonymous SNP and the second of which falls far from any known gene. What is the probability that the first SNP is truly associated with the trait, and how does this compare to the probability for the second? 

A potential answer to this question comes from the second line of research that motivates this work. In association studies where the phenotype being studied is gene expression (``eQTL" studies, for ``expression quantitative trait locus"), statistical models have been developed to identify shared characteristics of SNPs that influence gene expression \citep{Veyrieras:2008lk, Gaffney:2012vn, Lee:2009kx}. In a hierarchical modeling framework, the probability that a given SNP influences gene expression can then depend on these characteristics. The key fact that makes these models useful in the context of eQTL mapping is that there is a large number of unambiguous eQTLs in the genome on which a model can be trained. In the GWAS context, the number of loci unambiguously associated with a given trait has historically been very small; learning the shared properties of two or three loci is not a job well-suited to statistical modeling. However, large meta-analyses of GWAS now regularly identify tens to hundreds of independent loci that influence a trait (e.g. \cite{allen2010hundreds, teslovich2010biological}). The merits of hierarchical modeling in this context \citep{Lewinger:2007fk, heron2011exploration,Chen:2007kx} are thus worth revisiting. Indeed, Carbonetto and Stephens\cite{carbonetto2012integrated} have reported success in identifying loci involved in autoimmune diseases using a hierarchical model that incorporates information about groups of genes known to interact in a pathway. 

In this paper we present a hierarchical model for jointly analyzing GWAS and genomic annotations. We applied this model to GWAS of 18 diseases and traits; for each trait, we learned the relevant types of genomic information from a set of 450 genome annotations.

\section{Results}
We assembled a set of 18 GWAS with publicly available summary statistics and a large number (at least around 20) of loci associated with the trait of interest. These included studies of red blood cell traits \citep{van2012seventy}, platelet traits \citep{gieger2011new}, Crohn's disease \citep{jostins2012host}, BMI \citep{speliotes2010association}, lipid levels \citep{teslovich2010biological}, height \citep{allen2010hundreds}, bone mineral density \citep{estrada2012genome} and fasting glucose levels \citep{manning2012genome}. We used ImpG \citep{pasaniuc2013fast} to impute the summary statistics from each study for all common SNPs identified in European populations by the 1000 Genomes Project \citep{abecasis2010map}. Overall we successfully imputed association statistics for around 80\% of common SNPs (Supplementary Figure 1). We then assembled a set of genome annotations, paying specific attention to annotations available for many cell types since important regulatory elements may be active only in specific cell types. The main sources of genome annotations were 402 maps of DNase-I hypersensitivity in a wide range of primary cell types and cell lines \citep{Maurano:2012vn, thurman2012accessible}. We also included as annotations  the output from ``genome segmentation" of the six main ENCODE cell lines \citep{hoffman2013integrative}; for each section of the genome in each cell line, Hoffman et al. \cite{hoffman2013integrative} report whether the histone modifications in the region are consistent with enhancer activity, transcription start sites, promoter-flanking regions, CTCF binding sites, or repressed chromatin. Finally, we included elements of gene structures (protein-coding exons and 3' and 5' UTRs). In total, we used data from 18 traits and 450 genomic annotations. 
 
For each trait, we set out to identify which of the 450 annotations (if any) were enriched in genetic variants influencing the trait. To do this, we developed a hierarchical model that learns the shared properties of loci influencing a trait. The full details of the model are presented in the Methods, but can be summarized briefly. Conceptually, we break the genome into large, non-overlapping blocks (with an average size of 2.5 Mb). Let the prior probability that any block $k$ contains an association be $\Pi_k$. \emph{If} there is an association in block $k$, then let the prior probability that any SNP $i$ is the causal SNP be $\pi_{ik}$. We allow both $\Pi_k$ and $\pi_{ik}$ to depend on annotations of the region and SNP, respectively, and estimate these quantities based on the patterns of enrichment across the whole genome. We tested this approach using simulations based on real data from a GWAS of height (Supplementary Material).

\begin{figure}
\begin{center}
\includegraphics[scale = 0.8]{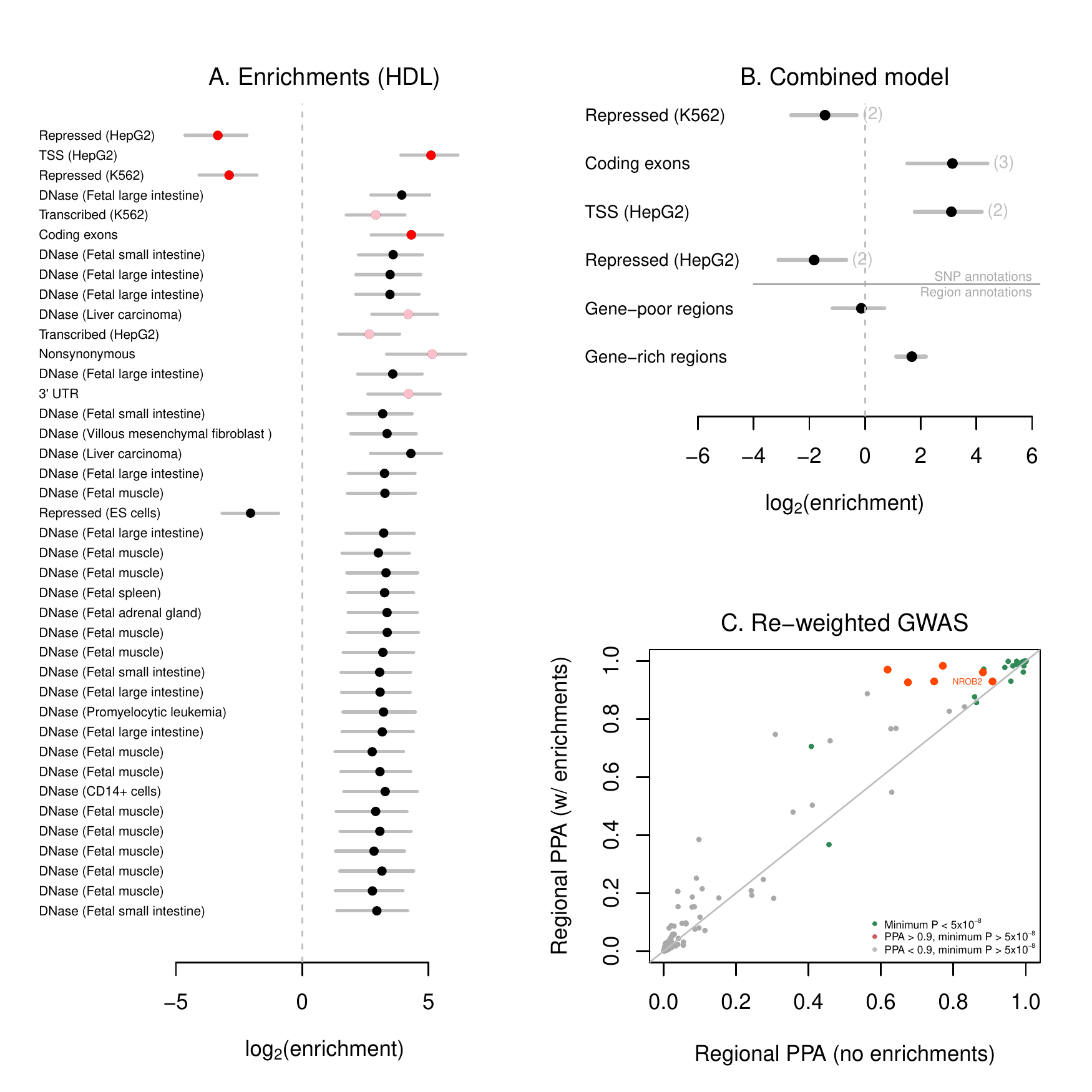}
\caption{. \textbf{Application of the model to HDL cholesterol. A. Single-annotation models.} We fit the model to each annotation individually, including a SNP-level effect for SNPs 0-5kb from a TSS, a SNP-level effect for SNPs 5-10kb from a TSS, a region-level effect for regions in the top third of gene density, and a region-level effect for regions in the bottom third of gene density. Plotted are the maximum likelihood estimates and 95\% confidence intervals of the enrichment parameter for each annotation. Annotations are ordered according to how much they improve the likelihood of the model (at the top are those that improve the likelihood the most). In red are the annotations included in the joint model, and in pink are the annotations that are statistically equivalent to those included in the combined model. \textbf{B. Joint model.} Using the algorithm described in the Methods we built a model combining multiple annotations. Shown are the maximum likelihood estimates and 95\% confidence intervals of the enrichment effects of each annotation. Note that though these are the maximum likelihood estimates, model choice was performed using a penalized likelihood. In parentheses next to each annotation (expect for those relating to distance to transcription start sites), we show the total number of annotations that are statistically equivalent to the included annotation in a conditional analysis. \textbf{C. Re-weighted GWAS.} We re-weighted the GWAS using the model with all the annotations in \textbf{B} (under the penalized enrichment parameters from Supplementary Table 9). Each point represents a region of the genome, and shown are the posterior probabilities of association (PPA) of the region in the models with and without the annotations.}\label{fig_mch}
\end{center}
\end{figure}

The methodology is best illustrated with an example. We started with an analysis of a GWAS of high-density lipoprotein (HDL) levels \citep{teslovich2010biological}. We first took each genomic annotation individually, and estimated its level of enrichment (or depletion) for loci that influence HDL (in the model, we additionally included a regional effect of gene density and and a SNP-level effect of distance to the nearest transcription start site, see the Methods for details). In Figure \ref{fig_mch}A, we show the top 40 annotations, ordered by how well each improves the fit of the model. Loci that influence HDL are most strongly enriched in enhancers identified in the HepG2 cell line, and most strongly depleted from genomic regions repressed in that same cell line. HepG2 cells are derived from a liver cancer; the relevance of this cell line to a lipid phenotype makes intuitive sense. However, there are many other additional (correlated) genome annotations that are enriched for loci that influence HDL (Figure \ref{fig_mch}A).  We thus built a model including multiple annotations; to mitigate over-fitting in this situation we used a cross-validation approach (Methods). The best-fitting model is shown in Figure \ref{fig_mch}B. It include both enhancers and repressed chromatin identified in HepG2 cells, as well as coding exons and chromatin repressed in K562 cells. Since many of the annotations are correlated, those included in the combined model are the ``best" representatives of sets of related annotations. We thus used conditional analysis to define the set of annotations represented by each member of the combined model (Methods; Figure \ref{fig_mch}A).

\begin{figure}
\begin{center}
\includegraphics[scale = 1]{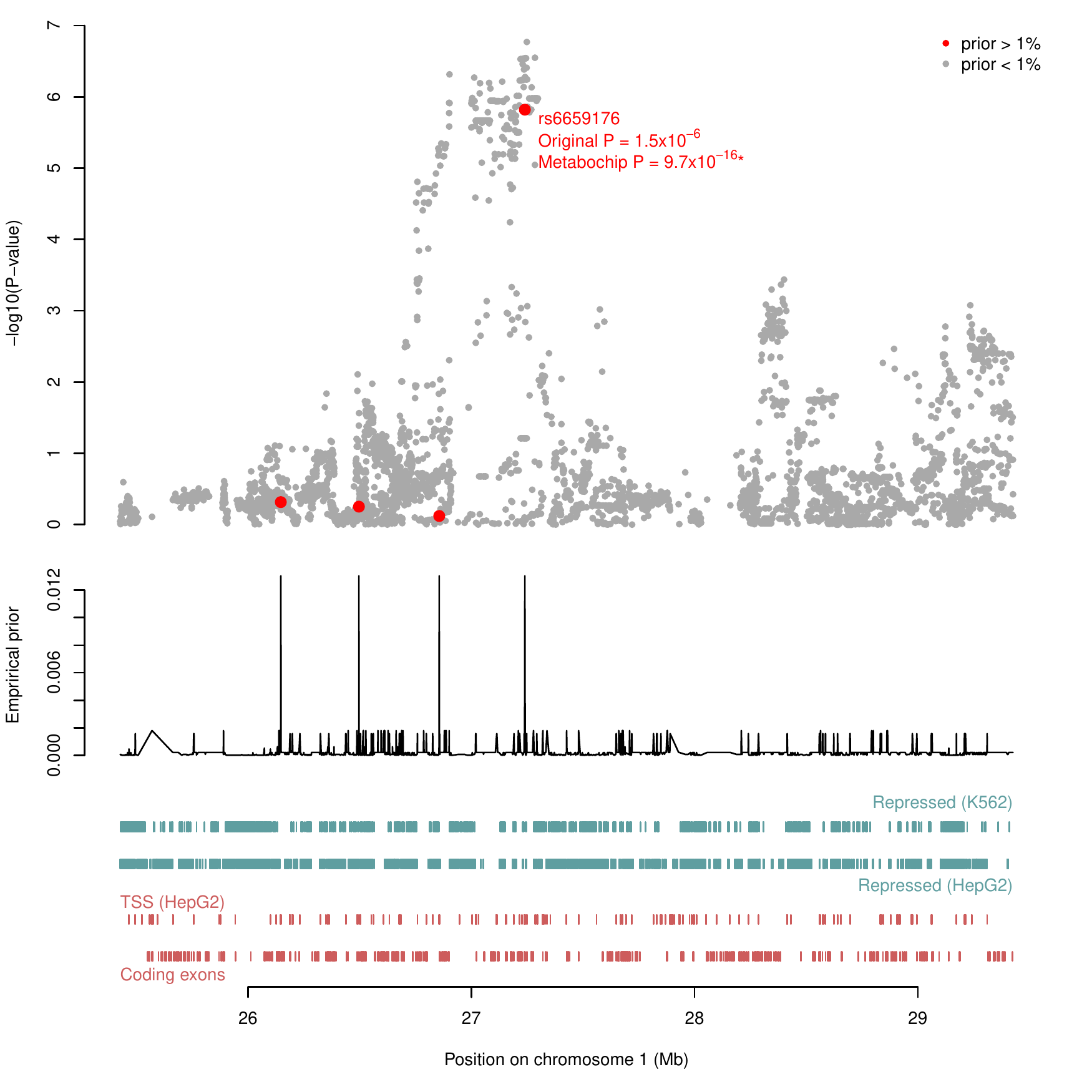}
\caption{. \textbf{Regional plot surrounding NR0B2.} In the top panel we plot the P-values for association with HDL levels at each SNP in this region. In the middle panel is the fitted empirical prior probability that each SNP is the causal one in the region, conditional on there being a single causal SNP in the region. This prior was estimated using the combined model with the annotations in Figure \ref{fig_mch}B. In the lower panel are the positions of the annotations included in the model.* The reported P-value is for rs12748152, which has $r^2 = 0.85$ with rs6659176.}\label{fig_mch_eif}
\end{center}
\end{figure}

A convenient side effect of fitting an explicit statistical model relating properties of SNPs to the probability of association is that we can use the functional information to re-weight the GWAS (Methods). We used the combined model for HDL to re-weight the association statistics across the genome (Figure \ref{fig_mch}C). There are several regions of the genome with strong evidence for association with HDL (posterior probability of association [PPA] over 0.9) only when using the model incorporating functional information. In Figure \ref{fig_mch_eif}, we show one such region, near the gene NR0B2. The model identifies the SNP rs6659176 as the most likely candidate to be the causal polymorphism in this region. This SNP has a P-value of $1.5 \times 10^{-6}$. However, this SNP falls falls in a coding exon (in fact it is nonsynonymous), leading the model to conclude that this P-value is in fact strong evidence for association. Indeed, larger studies of HDL have confirmed the evidence for association in this region (P-value of $9.7 \times 10^{-16}$ at rs12748152, which has $r^2 = 0.85$ with rs6659176  \citep{Global-Lipids-Genetics-Consortium:2013uq}). This region, though not this particular SNP, was also identified in a scan for SNPs influencing multiple lipid phenotypes \citep{Stephens:2013fk}.

\begin{figure}
\begin{center}
\includegraphics[scale = 0.6]{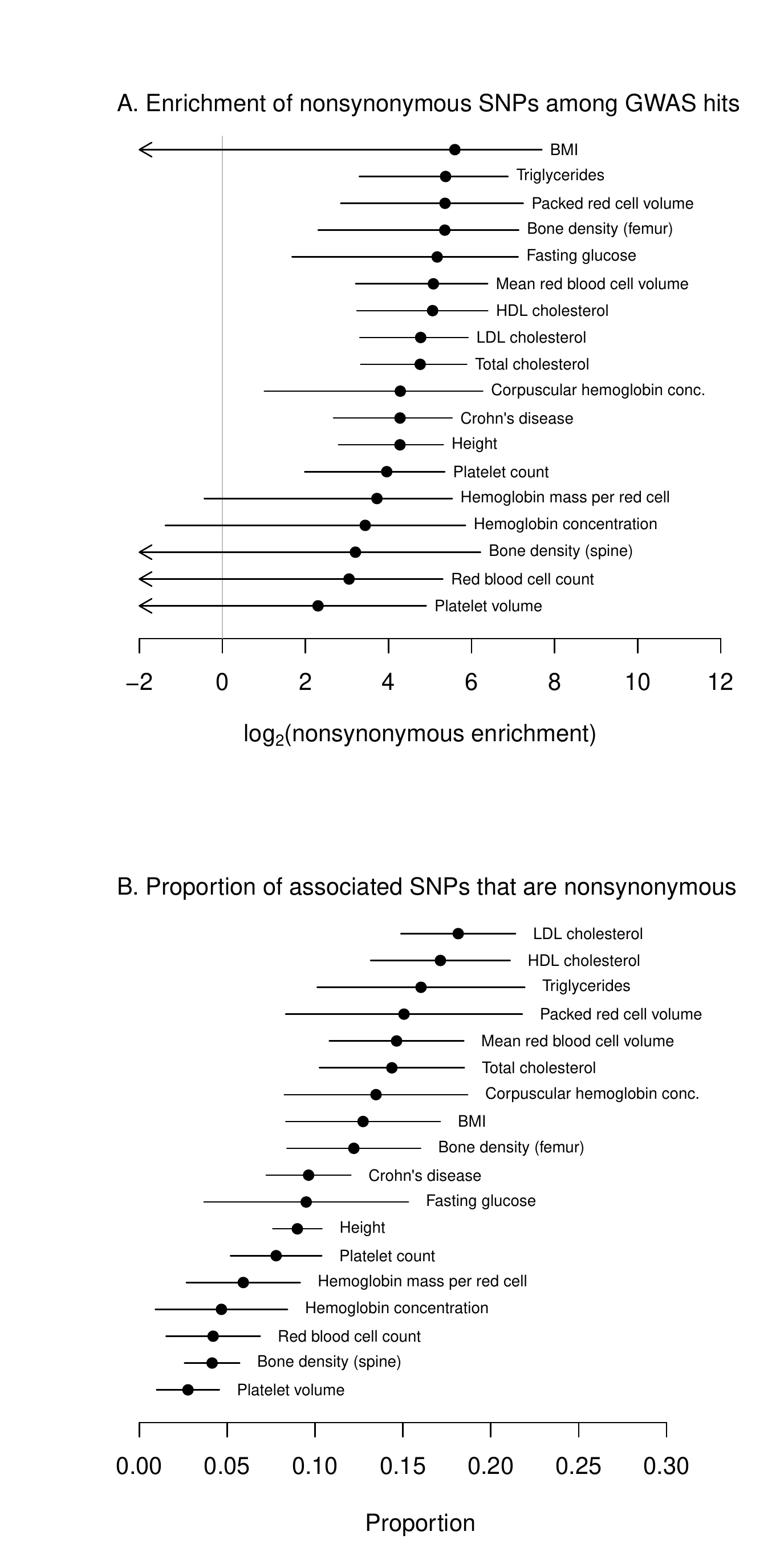}
\caption{. \textbf{Estimated role of protein-coding changes in each trait. A. Estimated enrichment of non-synonymous SNPs.} For each trait, we fit a model including an effect of non-synonymous SNPs and an effect of SNPs within 5kb of a TSS. Shown are the estimated enrichments parameters and 95\% confidence intervals for the non-synonymous SNPs. \textbf{B. Estimated proportion of GWAS hits driven by non-synonymous SNPs.} For each trait, using the model fit in \textbf{A.}, we estimated the proportion of GWAS signals driven by non-synonymous SNPs. Shown is this estimate and its standard error (Supplementary Material).}\label{fig_ns}
\end{center}
\end{figure}

We applied this method to all 18 traits. We were first interested in estimating the fraction of associations for each trait that can be explained by nonsynonymous polymorphisms versus polymorphisms that do not influence protein sequences. For each trait, we fit a model including promoters (SNPs within 5kb of a transcription start site) as well as nonsynonymous polymorphisms. For all traits, nonsynonymous polymorphsims are enriched among those that influence the trait, though this enrichment is not statistically significant for all traits (Figure \ref{fig_ns}A). This contrasts with synonymous polymorphisms, which are generally not enriched for polymorphisms that influence traits, with a few notable exceptions (like height and Crohn's disease, see Supplementary Figure 3). We then used these enrichments to estimate the fraction of associations for each trait that are driven by nonsynonymous polymorphisms (Supplementary Material). This fraction varies from around 2\% to around 20\%, with an average of 10\% (Figure \ref{fig_ns}B). We conclude that the relative importance of changes in protein sequence versus gene expression likely varies across traits.

We then used all 450 genome annotations to build models of enrichment for each trait. As for HDL, we first estimated enrichment levels individually for each annotation (Supplementary Figures 4-12). Clustering of phenotypes according to these enrichment levels recapitulated many known relationships between traits (Supplementary Figure 13). We then generated a combined model for each trait. The parameters of the combined models are shown in Figure \ref{fig_allpheno} and Supplementary Figure 14, and details of the exact annotations are in Supplementary Tables 2-20. In general, the models generated with this method are sparse and biologically interpretable. A few general patterns emerge from this analysis. Apart from the repeated occurrence of annotations related to protein-coding genes, marks of repressed chromatin are often significantly depleted for SNPs influencing traits. For example, SNPs influencing Crohn's disease are depleted from repressed chromatin identified in a lymphoblastoid cell line (Figure \ref{fig_allpheno}D; log$_2$ enrichment of -1.83, 95\% CI [-3.06, -0.78]), SNPs influencing height are significantly depleted from repressed chromatin in HeLa cells (Figure \ref{fig_allpheno}I; log$_2$ enrichment of -1.5, 95\% CI [-2.39, -0.71]), and SNPs influencing red blood cell volume are significantly depleted from repressed chromatin in an erythroblast-derived cell line (Figure \ref{fig_allpheno}F; log$_2$ enrichment of -3.91, 95\% CI [-6.25, -2.38]). 

We additionally observed cell type-specific enrichments in enhancer elements and DNase hypersensitive sites for SNPs that influence traits. Most of the observed enrichments are readily interpreted in light of the known biology of the trait. For example, SNPs that influence platelet volume and platelet count are enriched in open chromatin identified in CD34$^+$ cells, known to be on the cell lineage that leads to platelets \citep{Deutsch:2006fk} (Figure \ref{fig_allpheno}A,B; log$_2$ enrichment of 1.81, 95\% CI [0.59, 2.86] for platelet count; log$_2$ enrichment of 3.02, 95\% CI [1.69, 4.26] for platelet volume); and SNPs that influence corpuscular hemoglobin concentration are enriched in open chromatin identified in K562 cells, a cell line derived from a cancer of erythroblasts (Supplementary Figure 14E; $\log_2$ enrichment of 2.67, 95\% CI [0.61, 4.44]). For some traits, however, the connection between the trait and the tissues identified is not immediately obvious. For example, SNPs associated with platelet density are enriched in open chromatin in the spleen (Figure \ref{fig_allpheno}A; log$_2$ enrichment of 1.93, 95\% CI [0.59, 3.14]), and SNPs associated with height are enrichment in open chromatin in muscle (Figure \ref{fig_allpheno}I; log$_2$ enrichment of 2.27, 95\% CI [1.51, 3.02]; note that though there are all a large number of annotations equivalent to this annotation of open chromatin in fetal muscle, all of them are muscle-related; see Supplementary Figure 5B). 

\begin{figure}
\begin{center}
\includegraphics[scale = 1]{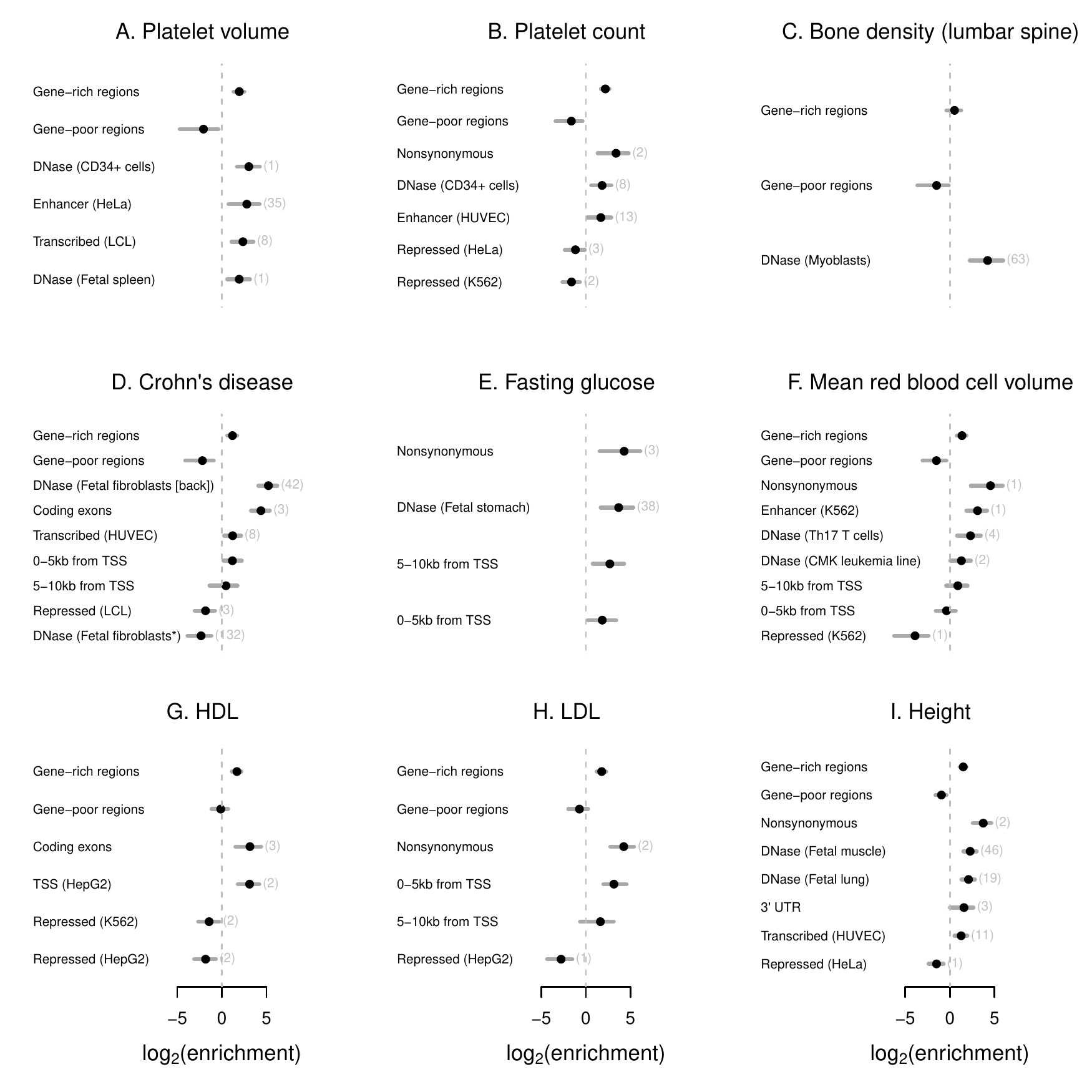}
\caption{. \textbf{Combined models for nine traits.} For each trait, we built a combined model of annotations using the algorithm presented in the Methods. Shown are the maximum likelihood estimates and 95\% confidence intervals for all annotations included in each model. Note that though these are the maximum likelihood estimates, model choice was done using a penalized likelihood (Methods). For the other nine traits, see Supplementary Figure 14. In parentheses next to each annotation (expect for those relating to distance to transcription start sites), we show the total number of annotations that are statistically equivalent to the included annotation in a conditional analysis (Methods).  *This annotation of DNase-I hypersensitive sites in fetal fibroblasts from the abdomen has a positive effect when treated alone; see the text for discussion.}\label{fig_allpheno}
\end{center}
\end{figure}

For two traits--Crohn's disease (Figure \ref{fig_allpheno}D) and red blood cell count (Supplementary Figure 14G)--we noticed that annotations initially identified as enriched for SNPs influencing the trait ended up in the combined model as being depleted for SNPs influencing the trait. On further examination (Supplementary Material), we found that these effects are due to statistical interactions. For example, when treated alone, SNPs that influence Crohn's disease are enriched in DNase-I hypersensitive sites identified in fetal fibroblasts from the abdomen (log$_2$ enrichment of 3.17, 95\% CI [1.66, 4.36]). However, DNase-I hypersensitive sites identified in fetal fibroblasts from the back show an even stronger enrichment (log$_2$ enrichment of 4.21, 95\% CI [2.99, 5.29]), and sites in common between the two annotation are intermediate (log$_2$ enrichment of 3.74, 95\% CI [2.29, 4.89]). This leads to an interaction where in the joint model the contribution of the DNase-I hypersensitive sites identified in fetal abdominal fibroblasts is negative. Though this is a statistical explanation for this observation, the biological explanation is not immediately clear. It seems likely that DNase-I hypersensitive sites are a heterogeneous set of different classes of elements, and that different experiments are more sensitive, for either technical or biological reasons, to subsets of these elements.

Finally, we explored the potential of this model to identify new loci (as in Figure \ref{fig_mch_eif}). In order to do this, one needs a threshold for ``significance" in this model, ideally with similar properties as the standard P-value threshold of $5\times 10^{-8}$. To calibrate the method, we used the fact that we initially applied our method to a study of four lipid traits that identified about 100 loci in a sample size of around 90,000 individuals \citep{teslovich2010biological}. Since then, larger studies have raised the number of loci associated with lipid traits to 157 \citep{Global-Lipids-Genetics-Consortium:2013uq}. If we treat a locus with a P-value of $5\times 10^{-8}$ in the larger study as a ``true positive" and a locus that does not reach this threshold in the larger study as a ``true negative", we can calibrate a threshold for posterior probability of association using the replication data (Supplementary Material). We found that a threshold of a regional PPA of 0.9 performed similarly to a stringent P-value threshold (Supplementary Figure 15). Combining the loci from both the standard P-value approach and our approach resulted in an approximately 5\% increase in the number of identified loci while still maintaining a false positive rate close to zero (Supplementary Figure 15, Supplementary Table 21). This is only a modest gain in power; that said, by applying this method to all 18 traits we identified 49 loci that did not reach a standard statistical significance threshold of $5\times 10^{-8}$ but have a PPA over 0.9 (Supplementary Table 22). Based on the above results for lipids, the level of evidence that these loci are true positives is approximately the same as those that have $P = 5\times 10^{-8}$ in a standard GWAS. Indeed, the majority of these loci have since been identified in larger cohorts than those used in this paper (Supplementary Table 22).

\section{Discussion}
In this paper, we have developed a statistical model for identifying genomic annotations that are most relevant to the biology of a given phenotype. We have shown that this model is able to scan through hundreds of genomic annotations to identify a sparse set of biologically-interpretable annotations without prior knowledge of the biology of the phenotype. 

\subsection*{Linking GWAS to biology}
Perhaps the most striking observation is that chromatin annotated as repressed in a given cell type is often depleted for SNPs that influence traits. Since approximately 60-70\% of the genome falls in this annotation in any given cell type (Supplementary Material), this information could dramatically limit the number of SNPs considered when fine-mapping loci identified in GWAS. Additionally, we identified several non-obvious connections between tissue and phenotypes. For example, SNPs that influence platelet volume are enriched in DNAse-I hypersensitive sites in the spleen (Figure 4A). Though the spleen functions in the removal of platelets from the bloodstream, the connection between its function and platelet volume is unclear.  An important next step will be to connect the identified non-coding variants in regulatory regions to changes in gene expression, which is presumably the mechanism by which they exert an effect on phenotypes.  Methods for inferring the casual chain between variation in DNAse-I sensitivity, variation in gene expression, and variation in phenotypes (e.g. \cite{giambartolomei2013bayesian, Degner:2012aa, Plagnol:2009aa, Nica:2010aa, Stephens:2013fk}) will be essential. 
%For example, SNPs that influence bone density in the spine and the femur are both enriched in DNAse-I hypersensitive sites identified in muscle-related tissues (myoblasts and the fetal heart; Figure \ref{fig_allpheno}D and Supplementary Figure 12B). Though it is well-known that muscle function correlates with bone mineral density, twin studies have suggested that this correlation is not due to genetic factors \citep{Arden:1997kx}. However, a modest shared genetic component between muscle function and bone mineral density cannot be excluded by these studies, and our results suggest that this possibility is worth re-examining. Additionally, we have shown that SNPs that influence Crohn's disease risk are enriched in DNase-I hypersensitive sites identified in fibroblasts (Figure \ref{fig_allpheno}D). Though Crohn's disease is an autoimmune disease, a major complication of the disease is intestinal stricture mediated through abnormal fibroblast activity \citep{Burke:2007vn}. Our results suggest that the intestinal response to inflammation, and not just the autoimmune activity causing inflammation,  may be modulated by genetic variation.

\subsection*{Modeling assumptions}
We have made several modeling assumptions that merit discussion. First, by splitting the genome into blocks based on numbers of SNPs, we are making the implicit assumption that the probability that a genomic region contains a SNP associated with a given phenotype depends on the SNP density rather than the physical size--that is, a short genomic region with a large number of SNPs is \emph{a priori} as likely to have an association as a long genomic region with few SNPs. We have also made a more restrictive assumption that there can be only a single causal SNP in a given genomic region. This assumption is a natural starting point, but as GWAS sample sizes increase even more it will begin to be untenable. Advances in methods for joint analysis of multiple SNPs (e.g. Yang et al. \cite{Yang:2012uq}) may provide a way forward in this situation. Finally, we note that the model is limited by the types of genomic annotations that are available, and the best annotations identified in the model may be ``proxies" for the truly-relevant annotations. For example, SNPs associated with height are enriched in DNase-I hypersensitive sites identified in the fetal lung (Figure \ref{fig_allpheno}I); taken literally, this would suggest that some SNPs influence height though lung development. An alternative possibility, however, is that patterns of open chromatin in the lung (which is of course a heterogeneous tissue) are useful proxies for patterns of open chromatin in a cell type that has not been profiled; this hypothetical cell type could in principle be present in any tissue.

\subsection*{Prospects for fine-mapping GWAS loci using functional genomic data}
We have primarily focused on using our model to identify annotations relevant to a trait of interest, though we have also explored using this information to identify novel loci. A third natural application, which we have not explored, is the possibility for fine-mapping GWAS loci using functional genomic information \citep{maller2012bayesian}. Indeed, the posterior probability that each SNP in a given genomic region is the causal one is explicitly included in our model. However, in current applications around 20\% of common SNPs are neither genotyped nor successfully imputed; this is a major limitation to fine-mapping that cannot be overcome with statistical means. As GWAS move to even denser genotyping or sequencing, we expect that re-visiting this issue will be fruitful.

\section*{Methods}
In this section we detail the specifics of the hierarchical model; a description of the data used is in the Supplementary Material. The model we propose is most closely related to that developed by Veyrieras et al. \cite{Veyrieras:2008lk} in the context of eQTL mapping. Conceptually, we split the genome into independent blocks, such that the blocks are larger than the extent of LD in the population. We then allow each block to contain either a single polymorphism that causally influences the trait or none. We model the prior probability that any given block contains an association and the conditional prior probability that any given SNP in the block is the causal one. The key is that we allow these probabilities to vary according to functional annotations--for example, gene-rich regions might be more likely to contain associations, and if there is an association, the causal polymorphism may be more likely to fall in a transcription factor binding site. We then estimate these priors using an empirical Bayes approach. Software implementing this model is available at \url{https://github.com/joepickrell/fgwas}. 

\subsection*{Computing the Bayes factor}

The basic building block of the model is a linear regression model. Consider a single SNP genotyped in $N$ phenotyped individuals.  Assume each individual has an associated measurement of a quantitative trait (we describe a slight modification for case-control studies later), and let $\vec y$ be the vector of phenotypes. Let $\vec g$ be the vector of genotypes (coded 0, 1, or 2 according to counts of an arbitrarily-defined allele). We use a standard additive linear model:

\begin{equation}
E[y_i] = \alpha + \beta g_i. 
\end{equation} 
We would like to compare two models: one where $\beta = 0$ and one where $\beta \ne 0$. A natural way to compare these two models is the Bayes factor:

\begin{equation} \label{BF_eq}
BF = \frac{\int P( \vec y | \vec g, H_1)}{\int P( \vec y | \vec g, H_0) }, 
\end{equation}  
\noindent where $H_1$ and $H_0$ represent the parameters of the the alternative and null models, respectively, and which are integrated out. 

To compute Equation \ref{BF_eq}, we use the approximate Bayes factor from Wakefield \cite{Wakefield:2008hc}. This Bayes factor has the practically important property that is can be calculated from a summary of the linear regression, without access to the underlying genotype vector $\vec g$. For completeness, we re-iterate here the underlying model. If $\hat \beta$ is the maximum likelihood estimator of $\beta$ and $\sqrt{V}$ is the standard error of $\hat \beta$,  Wakefield \cite{Wakefield:2008hc} suggests a model in which:

\begin{equation}
\hat \beta \sim N(\beta, V).
\end{equation}

Wakefield \cite{Wakefield:2008hc} places a normal prior on $\beta$, such that $\beta \sim N(0, W)$. Under this model Equation \ref{BF_eq} becomes:

\begin{equation} \label{eq_wakefield}
BF = \frac{\sqrt{1-r}}{\exp[-\frac{Z^2}{2}r]},
\end{equation} 
\noindent where $r = \frac{W}{V+W}$ and $Z = \frac{\hat \beta}{\sqrt{V}}$ (a standard Z-score). Thus, from a Z-score, an estimate of $V$, and the prior variance $W$, we can obtain a Bayes factor measuring the statistical support for a model in which a SNP is associated with a trait as compared to a model in which a SNP is not associated with a trait. Note, however, that because of linkage disequilibrium in the genome, any true causal association will lead to multiple true statistical associations. In all applications, we set $W = 0.1$ as the prior, such that the majority of the weight of the prior is on small effect sizes (results are robust to some variation in this prior; see Supplementary Material and Supplementary Figure 16).

\subsection*{Hierarchical model}
Now consider a set of $M$ SNPs, each of which has been genotyped in $N$ individuals in a GWAS. Our goal is to build a model to identify the shared characteristics of SNPs that causally influence a trait. Because of LD, there will be many associations in the genome that are not causal; however, these will all be restricted to a block around the truly causal site. We thus split the genome into contiguous blocks of size $K$ SNPs (in all of our applications, we set $K = 5,000$, though doubling this block size had little effect on the results; see Supplementary Material and Supplementary Figure 16), such that there are $M/K$ blocks. We choose the block size to be much larger than the extent of linkage disequilibrium in the population. Let $\Pi_k$ be the prior probability that block $k$ contains a causal SNP associated with the trait. The probability of the data (the set of observed phenotypes) is then:

\begin{equation}
P(\vec y) = \prod \limits_{k = 1}^{M/K} (1-\Pi_k) P^0_k + \Pi_k P^1_k,
\end{equation}  
\noindent where $P^0_k$ is the probability of the data in block $k$ under the model where there are no SNPs associated with the trait in the block, and $P^1_k$ is the probability of the data in block $k$ under the model where there is one SNP associated with the trait in the block. Further,

\begin{equation}
P^1_k = \sum \limits_{i \in S_k} \pi_{ik} P^1_{ik},
\end{equation}
\noindent where $S_k$ is the set of SNPs in block $k$, $\pi_{ik}$ is the prior probability that SNP $i$ is the causal SNP in the region conditional on there being an association in block $k$, and $P^1_{ik}$ is the probability of the data under the model where this SNP is associated with the trait. Note that this is not a multiple regression model where we jointly model the effects of multiple SNPs on a trait (as in, for example, Carbonetto and Stephens \cite{carbonetto2012integrated}). 

We can now allow the prior probabilities--both $\Pi_k$ (the prior on the block of SNPs containing an association) and $\pi_{ik}$ (the prior probability that SNP $i$ is the causal SNP assuming there is a single association in block $k$)--to depend on external information. We would also like to avoid subjective variation in $\Pi_k$ and $\pi_i$, but instead learn from the data itself which genomic annotations are most important. Specifically, we model the regional prior probability as:

\begin{equation}
\ln \bigg(\frac{\Pi_k}{1-\Pi_k}\bigg) = \kappa + \sum \limits_{l = 1}^{L_1} \gamma_{l} I_{kl},
\end{equation}
\noindent where $L_1$ is the number of region-level annotations in the model, $\gamma_l$ is the effect associated with annotation $l$ and $I_{kl}$ takes the value 1 if region $k$ is annotated with annotation $l$ and 0 otherwise. For example, in practice we will estimate a $\gamma$ parameter for regions of high or low gene density. We then model the SNP prior probability as:

\begin{equation}
\pi_{ik} = \frac{e^{x_i}}{\sum \limits_{j \in S_k} e^{x_j}},
\end{equation} 
\noindent where

\begin{equation} \label{annot_eq}
x_i = \sum \limits_{l = 1}^{L_2} \lambda_l I_{il},
\end{equation}
\noindent where $L_2$ is the number of SNP-level annotations in the model, $\lambda_l$ is the effect of SNP annotation $l$ and $I_{il}$ takes the value 1 if SNP $i$ falls in annotation $l$ and 0 otherwise. For example, in practice we will estimate a $\lambda$ parameter for nonsynonymous SNPs.

\subsection*{Fitting the model}
Combining terms above, we see that the likelihood of the data can be written down as:

\begin{align} \label{lk_eq}
L(\vec y | \theta) &= \prod \limits_ {k = 0}^{M/K} (1-\Pi_k) P^0_k + \Pi_k \sum \limits_{i = 0}^K \pi_{ij} P^1_{ik}\\
&= \prod \limits_{k = 0}^{M/K} P^0_k [  (1-\Pi_k) + \Pi_k \sum \limits_{i = 0}^{K} \pi_{ik} BF_i],
\end{align}
\noindent where $\theta$ is contains all the parameters of the model, most notably the set of annotation parameters. We maximize this function using the Nelder-Mead algorithm implemented in the GNU Scientific Library.

\subsection*{Shrinkage estimators of the annotation parameters}
While maximizing Equation \ref{lk_eq} gives the maximum likelihood estimates of all parameters, one concern is that there may be some level of overfitting. When comparing models, we instead shrink these parameters towards zero.  Specifically, we define a penalized log-likelihood function:

\begin{equation}
l^*(\vec y | \theta) = ln\bigg((L(\vec y | \theta)\bigg) - p \bigg(\sum \limits_{l = 1}^{L_1} \gamma_l^2 + \sum \limits_{l = 1}^{L_2} \lambda_l ^2\bigg).
\end{equation}
The penalty $p$ on the sum of the squared annotation parameters is the one used in ridge regression \citep{trevor2001elements}. In ridge regression, parameter estimates under this penalty are equivalent to estimating the posterior mean of the parameter if the prior distribution of the parameter is Gaussian \citep{trevor2001elements}; changing the tuning parameter $p$ is equivalent to changing the prior. We suspect that the interpretation in this model is similar. Since this penalized likelihood can not be used for formal statistical tests, we tune the $p$ parameter by cross-validation. An alternative approach here would be to explicitly put a prior on the enrichment parameters, but in the absence of a conjugate prior this would likely add substantially to the computational burden for little practical benefit. 

\subsection*{Cross-validation}
To compare models and tune the penalty $p$ in the penalized likelihood above, we used a 10-fold cross-validation approach. We split the chromosomal segments into 10 folds. Let $\theta^p_{-f}$ be the parameters of the model estimated while holding out the data from fold $f$ and under penalty $p$, and $l^{*}_{f}(\theta^p_{-f})$ be the penalized log-likelihood of the data in fold $f$ under the model optimized without using fold $f$. Then:

\begin{equation} \label{eq_xv}
l^{'}(\theta^p)  = \frac{1}{10} \sum \limits_{i = 1}^{10} l^{*}_{i}(\theta^p_{-i}).
\end{equation} 

Note that the size of the folds used in this cross-validation means that each fold excludes more than an entire chromosome. This means that no individual chromosome can have undue influence on the parameters included in the model.
\subsection*{Model choice}
Consider a single phenotype and a set of $L$ functional annotations of SNPs (in our case $L$ is in the hundreds). Including all $L$ SNP annotations in the model is neither biologically interesting nor computationally feasible. We thus set out to choose a relatively sparse model that fits the data.  We start with forward selection: for each of the $L$ annotations, we fit a model including a region-level parameter for regions in the top third of the distribution of gene density, a region-level parameter for regions in the bottom third of the distribution of gene density, a SNP-level parameter for SNPs from 0-5 kb from a TSS, a SNP-level parameter for SNPs from 5-10kb from a TSS, and a SNP-level parameter for the annotation in question. We then identify the set of annotations that significantly improve the model fit (as judged by the likelihood from Equation \ref{lk_eq}). We then:

\begin{enumerate}
\item Add the annotation that most significantly improves the likelihood to the model.
\item For each annotation identified as having a significant marginal effect, test a model including the annotation and those that have already been added.
\item If any annotation remains significant, go back to step 1.
\end{enumerate}

At this point, there are generally a small number of annotations in the model, but the model may be over-fit. We then switch to using the 10-fold cross-validation likelihood in Equation \ref{eq_xv}. We first tune the penalty parameter $p$ by finding the value of $p$ that maximizes the cross-validation likelihood. We then:
\begin{enumerate}
\item Drop each annotation from the model in turn, and evaluate the cross-validation likelihood. When dropping annotations, we additionally try dropping the region-level annotations on gene density and the SNP-level annotations on distance to the nearest TSS.
\item If a simpler model has a higher cross-validation likelihood than the full model, drop the annotation from the model and return to step 1.
\item Report the model that has the highest cross-validation likelihood.
\end{enumerate}

\subsection*{Approximating $V_i$}
In order to compute the Bayes factor in Equation \ref{eq_wakefield}, we need an estimate of $V_i$, the standard error of the estimated effect size of SNP $i$. In principle, this is trivial output from standard regression software; however, it is rarely reported. Instead, let $f_i$ be the minor allele frequency of SNP $i$ computed from an external sample of the same ancestry as the population in which the association study was done (we use data from the 1000 Genomes Project \citep{abecasis2010map}).  Let $N_i$ be the number of individuals in the association study at SNP $i$ (this can vary across SNPs due to missing data). Then:

\begin{equation}
V_i \approx \frac{1}{N_i f_i (1-f_i)}.
\end{equation}

\noindent Note that this variance is independent of the actual scale of the measurements; this is appropriate because the Z-scores are independent of the scale of the measurements as well. 

\subsection*{Case-control studies}
For all of the above, we have considered studies of quantitative traits. For a case-control study, we assume that we have summary statistics from logistic regression instead of linear regression. All aspects of the model are identical, with the exception of the approximation of $V_i$.  Define $N_{case}$ and $N_{control}$ as the numbers of cases and controls, respectively. Now, \citep{Wakefield:2008hc}:

\begin{equation}
V_i \approx  \frac{N_{case}+ N_{control}}{N_{case}N_{control} [ 2 f_i (1-f_i) + 4 f_i^2 - ( 2 f_i (1-f_i) + 4f_i^2)^2]}.
\end{equation}
\noindent The variance here is on a log-odds scale.

\subsection*{Posterior probabilities of association}
Once the model has been fit, we have empirical estimates of the prior probability that region $k$ contains an association, $\hat \Pi_k$ and the prior probability that SNP $i$ is the causal one, $\hat \pi_{ik}$ (conditional on there being an association). We define a Bayes factor summarizing the evidence for association in the \emph{region} (see, for example Maller et al. \cite{maller2012bayesian}):

\begin{equation}
BF^R_k = \sum_{i \in S_k} \hat \pi_{ik} BF_i,
\end{equation}
\noindent where $S_k$ is the set of SNPs in region $k$ and $BF_i$ is the Bayes factor for SNP $i$ (Equation \ref{eq_wakefield}). The posterior probability that region $k$ contains an association is then:

\begin{equation}
PPA^R_k = \frac{\hat \Pi_k BF^R_k/ (1- \hat \Pi_k)}{1+\hat \Pi_k BF^R_k/ (1- \hat \Pi_k)}
\end{equation}

We can also define the posterior probability that any given SNP $i$ in region $k$ is the causal one under our model:

\begin{equation}
PPA_{ik} = \frac{\hat \pi_{ik} BF_i}{\sum \limits_{j \in S_k} \hat \pi_{jk} BF_j}.
\end{equation}

\noindent This is similar to the calculation in Maller et al. \cite{maller2012bayesian}, except that we allow the prior probability $\pi_{ik}$ to vary across SNPs.  

Finally, we can define the posterior probability that any given SNP is causal. This is the posterior probability that the region contains a causal SNP times the posterior probability that the SNP is causal conditional on there being an association in the region. If SNP $i$ falls in region $k$:

\begin{equation}
PPA_{i} = PPA^{R}_k PPA_{ik}.
\end{equation}

\subsection*{Conditional analysis}
Because many of the annotations we consider are correlated, those ultimately included in the combined model for each trait (Figure \ref{fig_allpheno}) may be representatives of a large group of correlated annotations. For biological interpretation of the model, it is thus important to know which of the other annotations are interchangeable with those included in the model. 

To test this, we took an approach of conditional analysis. Consider two SNP-level annotations, with annotation parameters $\lambda_1$ and $\lambda_2$, respectively. In a joint model, we would jointly estimate both $\lambda_1$ and $\lambda_2$. However, we are interested in whether the second annotation adds information above and beyond that provided by the first annotation. We thus first estimate $\lambda_1$, then \emph{fix} this parameter to its maximum likelihood value $\hat \lambda_1$. We then estimate $\lambda_2 | \hat \lambda_1$ -- that is, we obtain the maximum likelihood estimate and 95\% confidence interval of $\lambda_2$ \emph{conditional} on a fixed value of $\lambda_1$. If this confidence interval does not overlap zero, this is evidence that the second annotation adds information to the model above that provided by the first. 

In practice, we first fit the combined model as described in the section ``Model choice" above. We then returned to the set of annotations that had significant marginal associations. For each annotation in the combined model, we took each of the other annotations in turn and tested whether the included annotation was significantly more informative than the non-included annotation. In Figure \ref{fig_allpheno} and Supplementary Figure 12, we display the total number of annotations represented by each one that is included in the combined model.
\clearpage
 
\paragraph{Acknowledgements.} We thank David Reich, Nick Patterson, Alkes Price, and Po-Ru Loh for helpful discussions and suggestions; and Jonathan Pritchard, Graham Coop, and two anonymous reviewers for comments on a previous version of this manuscript. We thank Nicole Soranzo for providing access to the platelet studies, and Luke Jostins for assistance in obtaining the Crohn's disease data. This work was supported by NIH postdoctoral fellowship GM103098 to JKP. 

\bibliography{../../Bib/bib}
\end{document}

\title{Supplementary material for: Joint analysis of functional genomic data and genome-wide association studies of 18 human traits}
\author{Joseph K. Pickrell$^{1,2}$\\
\small $^1$ New York Genome Center, New York, NY\\
\small $^2$ Department of Biological Sciences, Columbia University, New York, NY \\
\small Correspondence to: \url{jkpickrell@nygenome.org}
}
\maketitle

%\tableofcontents
\pagebreak
\tableofcontents
\pagebreak
\section{GWAS data}
\subsection{GIANT data}
We downloaded summary statistics from large GWAS of height \citep{allen2010hundreds} and BMI \citep{speliotes2010association} from \url{http://www.broadinstitute.org/collaboration/giant/index.php/GIANT_consortium}. The height summary statistics consisted of 2,469,635 SNPs either directly genotyped or imputed in an average of 129,945 individuals. We removed all SNPs with a sample size of less than 120,000 individuals. The BMI summary statistics consisted of 2,471,516 summary statistics either directly genotyped or imputed in an average of 120,569 individuals. We removed all SNPs with a sample size of less than 110,000 individuals. We then imputed summary statistics at SNPs identified in the 1000 Genomes Project as described in Section \ref{imputation_sec}.

\subsection{GEFOS data}
We downloaded summary statistics from large GWAS of bone mineral density \citep{estrada2012genome} from \url{http://www.gefos.org/?q=content/data-release}. There are two traits in these data: bone density measured in the femoral neck and bone density measured in the lumbar spine. The femoral neck bone density GWAS consisted of 2,478,337 SNPs, and the lumbar spine bone density consisted of 2,468,080 SNPs. Because the sample size at each SNP was not reported, we used the overall study sample sizes of 32,961 and 31,800 as approximations of the sample size at each SNP, and imputed summary statistics as described in Section \ref{imputation_sec}.

\subsection{IIBDGC data}
We downloaded summary statistics from a large GWAS of Crohn's disease \citep{jostins2012host} from \url{http://www.ibdgenetics.org/downloads.html}. The downloaded data consisted of 953,242 SNPs. Because the sample size at each SNP was not reported, we used the overall study sample sizes of 6,299 cases and 15,148 controls as approximations of the sample size at each SNP, and imputed summary statistics as described in Section \ref{imputation_sec}. Note that summary statistics from a GWAS of ulcerative colitis were also available from this site; however, these data contain a number of false positive associations that were filtered by  \citet{jostins2012host} using criteria that were not available to us. We thus only used the Crohn's disease association study.

\subsection{MAGIC data}
We downloaded summary statistics from a large GWAS of fasting glucose levels \citep{manning2012genome} from \url{http://www.magicinvestigators.org/downloads/}. The downloaded data consisted of  2,628,880 SNPs. Because the sample size at each SNP was not reported, we used the overall study sample size of 58,074 as an approximation of the sample size at each SNP, and imputed summary statistics as described in Section \ref{imputation_sec}.

\subsection{Global lipid genetics consortium data}
We downloaded summary statistics from a large GWAS of lipid traits \citep{teslovich2010biological} from \url{http://www.sph.umich.edu/csg/abecasis/public/lipids2010/}. These data consist of summary statistics for association studies of four traits: LDL cholesterol, HDL cholesterol, trigylcerides, and total cholesterol. The HDL data consisted of  2,692,429 SNPs genotyped or imputed in an average of 88,754 individuals, the LDL data consisted of 2,692,564 SNPs genotyped or imputed in an average of 84,685 individuals, the total cholesterol data consisted of 2,692,413 SNPs genotyped or imputed in an average of 89,005 individuals, and the triglycerides data consisted of  2,692,560 SNPs genotypes or imputed in an average of 85,691 individuals. For all traits, we removed SNPs with a sample size less than 80,000 individuals, and imputed summary statistics as described in Section \ref{imputation_sec}.

To calibrate significance thresholds, we additionally used summary statistics from \citet{Global-Lipids-Genetics-Consortium:2013uq}. These were downloaded from  \url{http://www.sph.umich.edu/csg/abecasis/public/lipids2013/}.

\subsection{Red blood cell trait data}
We obtained summary statistics from a large GWAS of red blood cell traits \citep{van2012seventy} from the European Genome-Phenome Archive (accession number EGAS00000000132). We downloaded summary statistics from association studies of six traits: hemoglobin levels,  mean cell hemoglobin (MCH), mean corpuscular hemoglobin concentration (MCHC), mean cell volume (MCV), packed cell volume (PCV), and red blood cell count (RBC). The hemoglobin level data consisted of 2,593,078 SNPs genotyped or imputed in 50,709 individuals, the MCH data consisted of 2,586,785 SNPs genotyped or imputed in an average of 43,127 individuals, the MCHC data consisted of 2,588,875 SNPs genotyped or imputed in an average of 46,469 individuals, the MCV data consisted of 2,591,132 SNPs genotyped or imputed in an average of 47,965 individuals, the PCV data consisted of 2,591,079 SNPs genotyped or imputed in an average of 44,485 individuals, and the RBC data consisted of 2,589,454 SNPs genotyped or imputed in an average of 44,851 individuals. We removed all SNPs with a sample size of less than 50,000 individuals (for hemoglobin levels) or 40,000 individuals (for the other traits), and imputed summary statistics as described in Section \ref{imputation_sec}.

\subsection{Platelet traits}
Summary statistics from a large GWAS of platelet traits \citep{gieger2011new} were generously provided to us by Nicole Soranzo. The data consist of summary statistics from association studies of two traits: platelet counts and mean platelet volume. The platelet count data consisted of 2,705,636 SNPs genotyped or imputed in an average of 44,217 individuals, and the platelet volume data consisted of 2,690,858 SNPs genotyped or imputed in an average of 16,745 individuals. We removed all SNPs with sample sizes less than 40,000 (for platelet counts) or 15,000 (for platelet volume), and imputed summary statistics as described in Section \ref{imputation_sec}. 

\section{Functional genomic data}
\subsection{DNase-I hypersensitivity data}
We downloaded DNase-I hypersensitivity data from two sources. The first was a set of regions defined as DNase-I hypersensitive by \citet{Maurano:2012vn} in 349 samples. We downloaded .bed files for 349 samples from \url{http://www.uwencode.org/proj/Science_Maurano_Humbert_et_al/} on February 13, 2013. These samples include 116 samples from cell lines or sorted blood cells, and 333 samples from primary fetal tissues. These latter samples were sampled from several tissues at various time points; we treated each track as independent rather than pooling data from tissues, since different experiments may have slightly different properties. The tissues in this latter group are fetal heart, fetal brain, fetal lung, fetal kidney, fetal intestine (large and small), fetal muscle, fetal placenta, and fetal skin. 

The second was a set of regions defined as DNase-I hypersensitive by the Crawford lab in the context of the ENCODE project \citep{thurman2012accessible}. We downloaded .bed files for 53 samples from \url{http://ftp.ebi.ac.uk/pub/databases/ensembl/encode/integration_data_jan2011/byDataType/openchrom/jan2011/fdrPeaks/} on March 29, 2013. We restricted ourselves to the files labeled as being generated at Duke University. Each experiment defined a set of regions of open chromatin in a particular cell type or cell line.

The ``Duke" DNase-I hypersensitive sites are all of exactly 150 bases in length, and each annotation covers approximately 1\% of the genome (range: 0.4 - 1.9 \% of the genome). The ``Maurano" DNase-I hypersensitive sites are on average 514 bases long, and each covers on average 2.7\% of the genome (range: 0.9-5.1 \% of the genome). 

\subsection{Chromatin state data}
We downloaded the ``genome segmentations" of the six ENCODE cell lines \citep{hoffman2013integrative} from \url{http://ftp.ebi.ac.uk/pub/databases/ensembl/encode/integration_data_jan2011/byDataType/segmentations/jan2011/hub/} on December 18, 2012. We used the ``combined" segmentation from two algorithms. This segmentation splits the genome into non-overlapping regions described as CTCF binding sites, enhancers, promoter-flanking regions, repressed chromatin, transcribed regions, transcription start sites, and weak enhancers. This segmentation was done independently in each of six cell lines, for a total of 42 annotations. 

Overall the ``repressed chromatin" mark covers the largest fraction of the genome, on average 66\% (ranging from 60\% for HUVEC cells to 70\% for H1 ES cells). The ``transcribed" mark covers on average 13\% of the genome, the ``CTCF" mark 1\% of the genome, the ``enhancer" mark 0.9\% of the genome, the ``TSS" mark 0.7\% of the genome, the ``weak enhancer" mark 0.4\% of the genome, and the ``promoter-flanking" mark 0.2\% of the genome. The remainder of the genome is not mappable by short reads and it thus excluded from these annotations.

\subsection{Gene models}
We downloaded the Ensembl gene annotations from the UCSC genome browser on May 21. Annotations of nonsynonymous and synonymous status for all SNPs in phase 1 of the 1000 Genomes Project were obtained from \url{ftp://ftp-trace.ncbi.nih.gov/1000genomes/ftp/phase1/analysis_results/functional_annotation/annotated_vcfs/}. Coding exons cover about 3\% of the genome, while 3' UTRs and 5' UTRs cover 2\% and 0.6\% of the genome, respectively. 

\section{Imputation of summary statistics} \label{imputation_sec}
\begin{figure}
\begin{center}
\includegraphics[scale = 0.8]{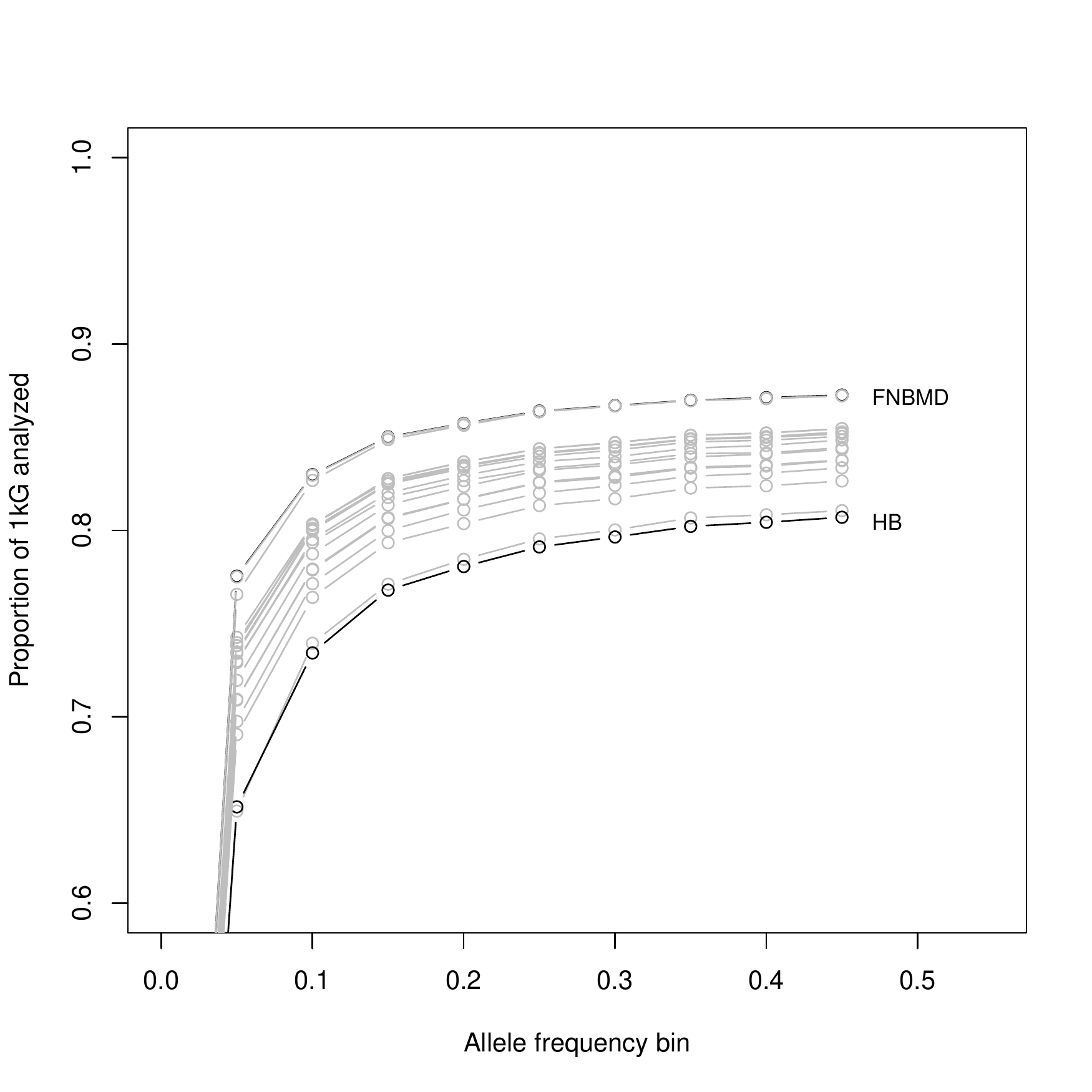}
\caption{. \textbf{Proportion of SNPs in the 1000 Genomes Project either genotyped or successfully imputed.} For each trait, we split all SNPs in phase 1 of the 1000 Genomes Project into bins based on their minor allele frequency in the European population. Bin sizes were of 5\% frequency. Shown are the proportions of SNPs in each bin that were either genotyped or successfully imputed for each trait (the points are at the lower ends of the bins, such that the point at 45\% frequency contains all SNPs from 45\%-50\% minor allele frequency). Labeled are the traits with the lowest and highest coverage. HB = hemogobin levels, FNBMD = femoral neck bone mineral density.}\label{fig_imp}
\end{center}
\end{figure}

We used ImpG v1.0 \citep{pasaniuc2013fast} under the default settings to impute summary statistics from all GWAS. As a reference panel, we used all haplotypes from European individuals in phase 1 of the 1000 Genomes Project, and only used SNPs with a minor allele frequency greater than 2\%. The reference haplotype files were derived from the 1000 Genomes integrated phase 1 v3.20101123 calls, downloaded from \url{ftp://ftp-trace.ncbi.nih.gov/1000genomes/ftp/phase1/analysis_results/integrated_call_sets/}. We used all 379 individuals labeled as ``European". After imputation, we removed all imputed SNPs with a predicted accuracy (in terms of correlation with the true summary statistics) less than 0.8. Overall, for each GWAS, we successfully imputed about 75-80\% of SNPs with a minor allele frequency over 10\% (Figure \ref{fig_imp}). 

To verify that imputation did not induce inflation of the test statistics, we computed the genomic control inflation factor $\lambda_{GC}$ \citep{bacanu2002association} before and after imputation (Supplementary Table \ref{lambda_table}). In all studies, inflation decreased after imputation, sometimes leading to a marked deflation in the test statistics. This is consistent with previous observations using this software \citep{pasaniuc2013fast}. The reason for this deflation is the shrinkage prior used in the imputation, which leads to conservative estimates of significance (imposed to strictly avoid false positive associations).

\section{Details of application of the hierarchical model}
\subsection{Simulations}
To test the performance of the model, we performed simulations using a GWAS of height \citep{allen2010hundreds}. Using the imputed summary statistics, we split the genome into blocks of 5,000 SNPs, then extracted the blocks with a genome-wide significant SNP reported in \citet{allen2010hundreds}. In each block, we had a reported Z-score for each SNP. To simulate annotations, we called the SNP with the smallest P-value in the region the ``causal" SNP. We then simulated annotations by placing all non-``casual" SNPs in an annotation with rate $r_1$, and all ``casual" SNPs in the annotation with rate $r_2$. We also varied the numbers of blocks included in the model. In each simulation, we randomly assigned SNPs to annotations according to determined rates, then ran our model under the assumption that $\Pi_k = 1$, that is, all blocks contain a causal SNP. We then calculated power as the fraction of simulations in which the confidence intervals of the annotation effect did not overlap zero. 

\begin{figure}
\begin{center}
\includegraphics[scale = 0.8]{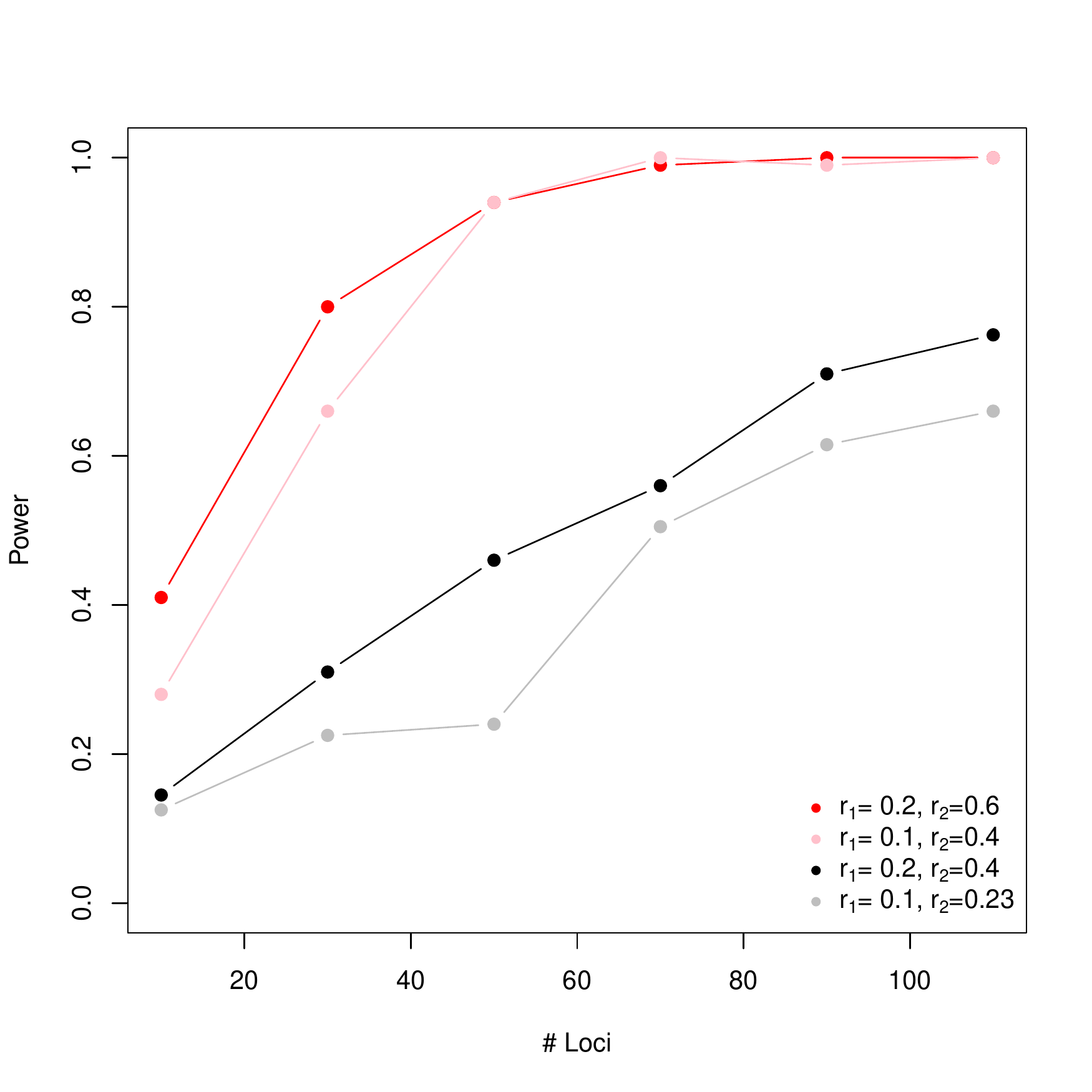}
\caption{. \textbf{Power to detect a significant annotation.} We simulated GWAS data under different levels of enrichment of causal SNPs in an annotation (see Supplementary Text), then evaluated the power of the method to detect the enrichment with different numbers of loci. In red and pink are log$_2$-enrichments of 2.6, and in black and grey are log$_2$-enrichments of 1.4.}\label{fig_power}
\end{center}
\end{figure}

We chose parameter settings of $r_1$ and $r_2$ such that the enrichment factors were similar to those in observed data (log-enrichment of 0.98 and 1.80). We chose $r_1$ to be either 0.2 and 0.1. For each set of parameters, we simulated 100 annotations and ran the model separately on each. Shown in Figure \ref{fig_power} is the power of the model. As expected, power increased as $r_1$ or the effect size increased, and as the number of loci increased.

\subsection{Robustness to choice of prior and window size}
There are two parameters in the model that are set by the user--the prior variance $W$ on the effect size and the window size defining ``independent" blocks of the genome.  We empirically tested the robustness of the model to variation in these parameters using the Crohn's disease dataset. We ran the model on each annotation using $W = 0.1$ and $W= 0.5$, additionally including--as in our main analyses--region-level parameters for regions in the top third and bottom third of gene density and SNP-level parameters for SNPs located from 0-5kb from a transcription start site and SNPs 5-10kb from a transcription start site. Plotted in Figure \ref{fig_prior}A are these annotation parameter estimates for all annotations where the 95\% confidence intervals did not overlap 0 in at least one run. The estimates from the two runs with different priors are highly correlated. We additionally tested window sizes of 5,000 SNPs and 10,000 SNPs (both with $W = 0.1$). The annotation effect estimates from these two window sizes are plotted in Figure \ref{fig_prior}B, and again are highly correlated.

\subsection{Quantifying the relative roles of coding versus non-coding changes in each phenotype}
To generate Figure 3 in the main text, we fit a model to each GWAS where we included region-level annotations for regions in the top third and bottom third of the distribution of gene density, and SNP-level annotations for non-synonymous SNPs and SNPs within 5kb of a transcription start site. Shown in Figure 3A in the main text are the estimates of the enrichment parameter for non-synonymous SNPs. At each SNP, the result of this model is the posterior probability that the SNP is casual (see Equation 19 in the main text). If we let this posterior probability at SNP $i$ be $PPA_i$, then the fraction of causal SNPs that are non-synonymous, $f_{NS}$ is:

\begin{equation}
f_{NS} = \frac{\sum_i PPA_i I^{NS}_i}{\sum_i PPA_i},
\end{equation}
\noindent where $I^{NS}_i$ is an indicator variable that takes value one if SNP $i$ is non-synonymous and zero otherwise. To get error bars on this fraction, we performed a block jackknife. We split the genome into 20 blocks with equal numbers of SNPs. If $f^j_{NS}$ is the estimate of the fraction of casual SNPs that are non-synonymous excluding block $j$, then:

\begin{equation}
SE = \sqrt{ \frac{19}{20} \sum \limits_{j = 1}^{20} (f^j_{NS} - \bar f_{NS})^2  },
\end{equation}
\noindent where $\bar f_{NS} = \frac{1}{20}\sum \limits_{i = 1}^{20} f^i_{NS}$. In Supplementary Figure \ref{fig_syn}, we show the corresponding results for synonymous SNPs.

\subsection{Interaction effects in annotation models}
As noted in the main text, there were two cases in which the sign of the annotation effect flipped between the single annotation models and the combined models. These were Crohn's disease (Supplementary Table \ref{cd_table}) and red blood cell count (Supplementary Table \ref{rbc_table}). In the main text we discuss the Crohn's disease example. For the red blood cell count example, note that SNPs influencing this trait are enriched in the annotation of DNAse-I hypersensitive sites in the fetal renal pelvis when this annotation is considered alone (log$_2$ enrichment of 2.48, 95\% CI [0.04, 4.17]). This annotation is correlated with the fetal stomach annotation, which has a log$_2$ enrichment of 4.83 (95\% CI [3.30, 6.45]) when treated alone. The SNPs in both of these annotations have a log$_2$ enrichment of 2.41 (95\% CI [-1.83, 4.23]), which leads to the interaction effect. Essentially the signal in the fetal stomach is driven by those SNPs that fall in DNase-I hypersensitive sites in the fetal stomach but \emph{not} the fetal renal pelvis. This suggests that there are a subset of DNase-I hypersensitive sites that are of particular interest for this phenotype. The interpretation of the Crohn's disease example is similar.

\subsection{Calibrating a ``significance" threshold}
For each genomic region, our method estimates the posterior probability that the region contains a SNP associated with a trait. If the model were a perfect description of reality, this probability could be interpreted literally. Since the model is not perfect, however, we sought a more empirical calibration. We used the fact that we initially ran the method on the GWAS data reported by \citet{teslovich2010biological} on four lipid traits. Since then, a GWAS with more individuals (though at a considerably smaller number of SNPs) has been reported for these four traits \citep{Global-Lipids-Genetics-Consortium:2013uq}. This latter study contains many of the individuals from the former (which had approximately 90,000 individuals), as well as about 80,000 more individuals. However, the additional individuals were genotyped in the Metabochip \citep{Voight:2012fk}, which has less than 200,000 markers, rather than the more dense standard GWAS arrays. This means that some regions of the genome do not benefit from the larger sample size.

For each region of the genome for each of the four traits, we built a table containing the minimum P-value from \citet{teslovich2010biological}, the posterior probability of association in the region (computed using the data from \citet{teslovich2010biological}), the minimum P-value from \citet{Global-Lipids-Genetics-Consortium:2013uq}, and the sample size used to get this minimum P-value (from \citet{Global-Lipids-Genetics-Consortium:2013uq}). We discarded regions where sample size at the SNP with the minimum P-value in the replication data set was smaller than 120,000 (since in these regions there is essentially no new data). We then coded each region as a ``true positive" if the minimum P-value from \citet{Global-Lipids-Genetics-Consortium:2013uq} was less than $5\times 10^{-8}$ and a ``true negative" otherwise. In Figure \ref{fig_roc}, we plot the number of ``true positives" and ``false negatives" that exceed various P-value and PPA thresholds. Note that since the data in \citet{Global-Lipids-Genetics-Consortium:2013uq} is not independent of that in \citet{teslovich2010biological}, this comparison is not appropriate for evaluating the relative performance of P-values versus the PPA. Our goal was simply to find a PPA threshold with similar performance in terms of reducing the number of false positives as the standard P-value threshold of $5\times 10^{-8}$. 

By visual inspection we set a PPA threshold at 0.9 (Figure \ref{fig_roc}). At this threshold, we identify 45 ``true positives" and zero ``false positives" for HDL, 43 and 1 for LDL, 47 and zero for total cholesterol, and 27 and zero for triglycerides. These are similar to the numbers for a P-value threshold of $5\times 10^{-8}$ (Supplementary Table \ref{roc_table}). Combining the loci identified by both methods leads to 48 loci for HDL  (versus 43 using a P-value threshold), 44 for LDL (versus 40), 51 for TC (versus 51) and 30 for TG (versus 29). This is on average an increase of 6\% in the number of loci identified. Note that this number is likely a lower bound, since the P-values in the replication study are naturally highly correlated to those in the initial study since they use many of the same individuals. A proper comparison would use a completely separate, large set of individuals to determine ``true positives" and ``true negatives", but such samples are not yet available.

\subsection{Identification of novel loci}
For each fitted model (using the parameters from Supplementary Tables \ref{bmi_table}-\ref{tg_table} estimated using the penalized likelihood), we calculated the posterior probability of association in each genomic region. We then identified all regions with a PPA greater than 0.9 but that had a minimum P-value less than $5\times10^{-8}$. For each remaining region, we identified the ``lead" SNP as the SNP with the largest posterior probability of being the causal SNP in the region. If this SNP was within 500kb of a SNP with $P < 5\times 10^{-8}$ (this can happen because we use non-overlapping windows and sometimes the best SNP is at the edge of the region), we removed it.  We also manually removed two regions (surrounding rs8076131 in Crohn's disease and surrounding rs11535944 in HDL), where the ``new" association was in LD with a previously reported SNP over 500kb away. In Supplementary Table \ref{assoc_table}, we show the remaining SNPs; these regions are high-confidence associations that did not reach traditional genome-wide significance. 

\begin{figure}
\begin{center}
\includegraphics[scale = 0.6]{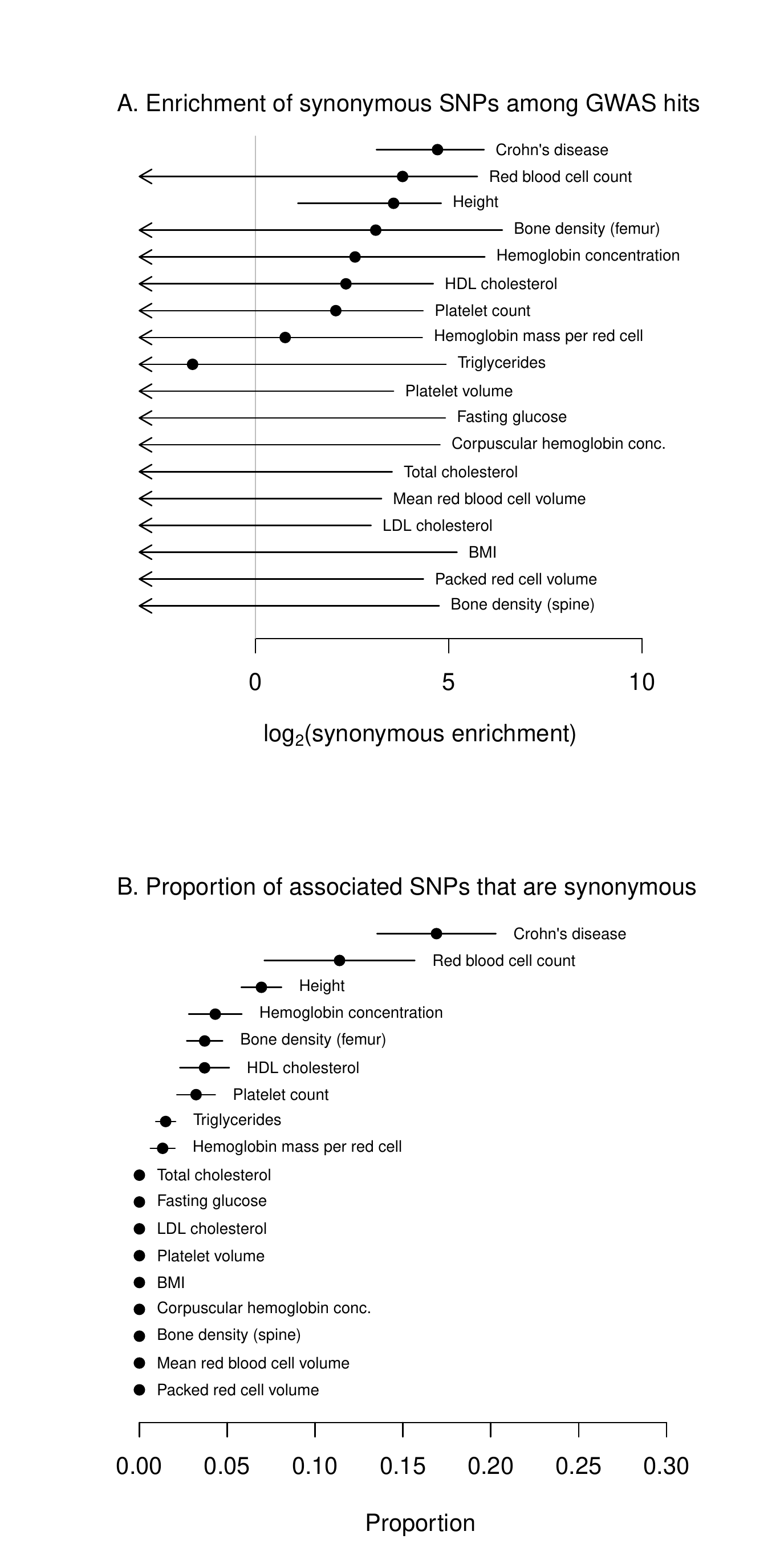}
\caption{. \textbf{Estimated role of synonymous polymorphisms in each trait. A. Estimated enrichment of synonymous SNPs.} For each trait, we fit a model including an effect of synonymous SNPs and an effect of SNPs within 5kb of a TSS. Shown are the estimated enrichments parameters and 95\% confidence intervals for the synonymous SNPs. \textbf{B. Estimated proportion of GWAS hits driven by synonymous SNPs.} For each trait, using the model fit in \textbf{A.}, we estimated the proportion of GWAS signals driven by synonymous SNPs. Shown is this estimate and its standard error.}\label{fig_syn}
\end{center}
\end{figure}

\begin{figure}
\begin{center}
\includegraphics[scale = 0.8]{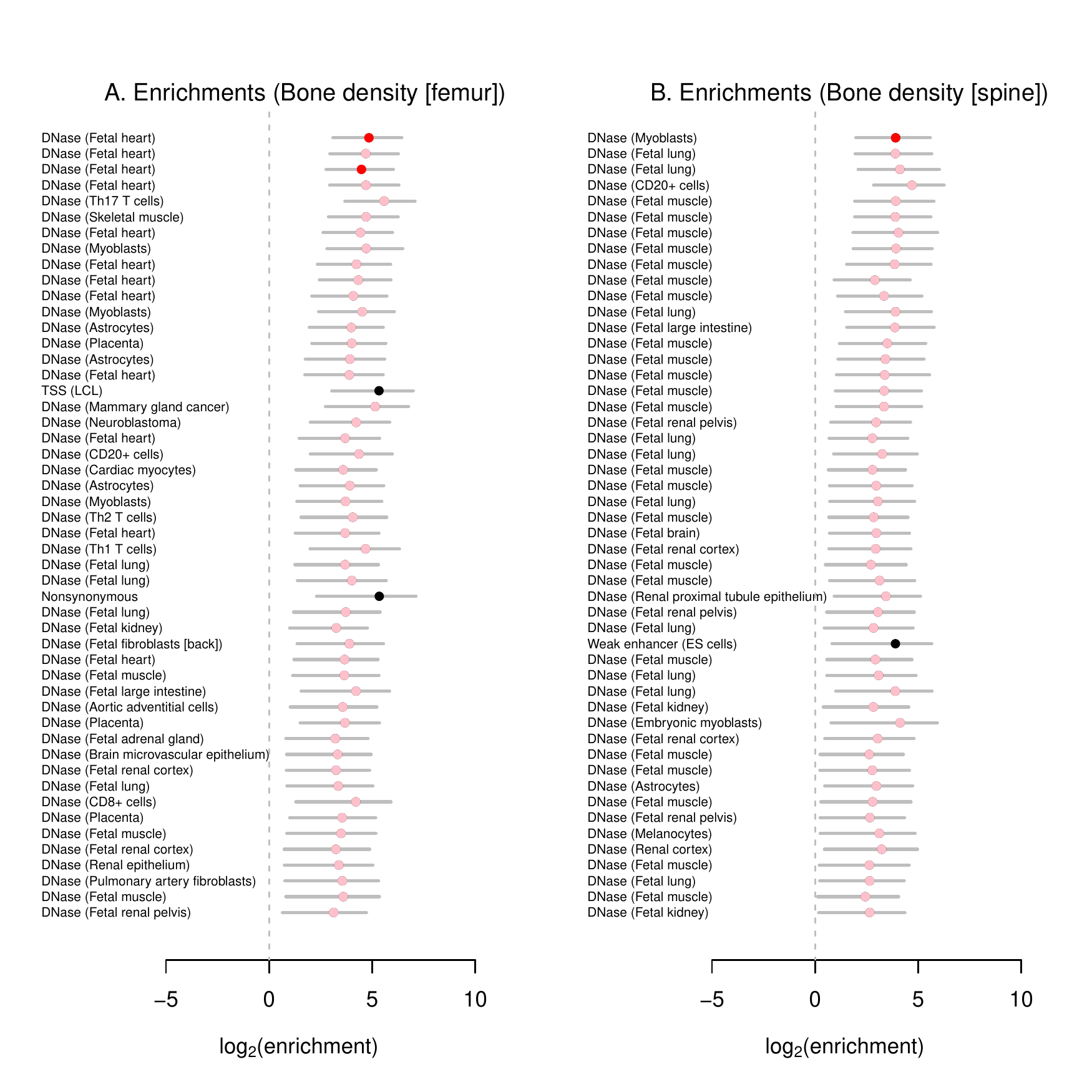}
\caption{. \textbf{Annotation effects in the bone mineral density data.} We estimated an enrichment parameter for each annotation individually in the GWAS for \textbf{A.} bone density in the femoral neck and \textbf{B.} bone density in the lumbar spine. Shown are the maximum likelihood estimates and 95\% confidence intervals. Annotations are ranked according to how much each improves the fit of the model; shown are the 50 annotations that most improve the model (or if there were less than 50 significant annotations, all of the significant annotations). In red are the annotations included in the combined model, and in pink are annotations that are statistically equivalent to those in the combined model.}\label{fig_bmd}
\end{center}
\end{figure}

\begin{figure}
\begin{center}
\includegraphics[scale = 0.8]{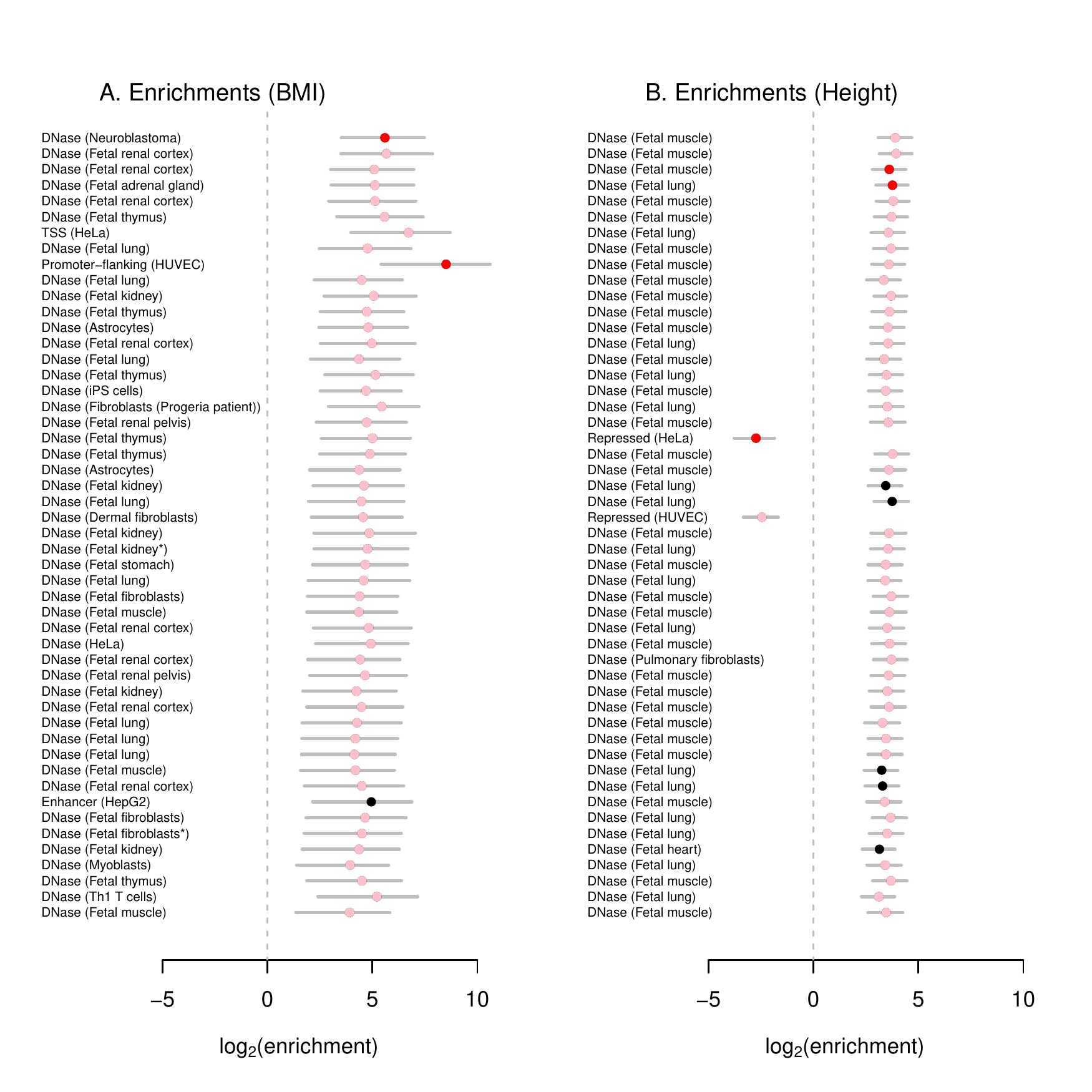}
\caption{. \textbf{Annotation effects in the GIANT data.} We estimated an enrichment parameter for each annotation individually in the GWAS for \textbf{A.} BMI and \textbf{B.} height. Shown are the maximum likelihood estimates and 95\% confidence intervals. Annotations are ranked according to how much each improves the fit of the model; shown are the 50 annotations that most improve the model (or if there were less than 50 significant annotations, all of the significant annotations). In red are the annotations included in the combined model, and in pink are annotations that are statistically equivalent to those in the combined model.}\label{fig_giant}
\end{center}
\end{figure}

\begin{figure}
\begin{center}
\includegraphics[scale = 0.8]{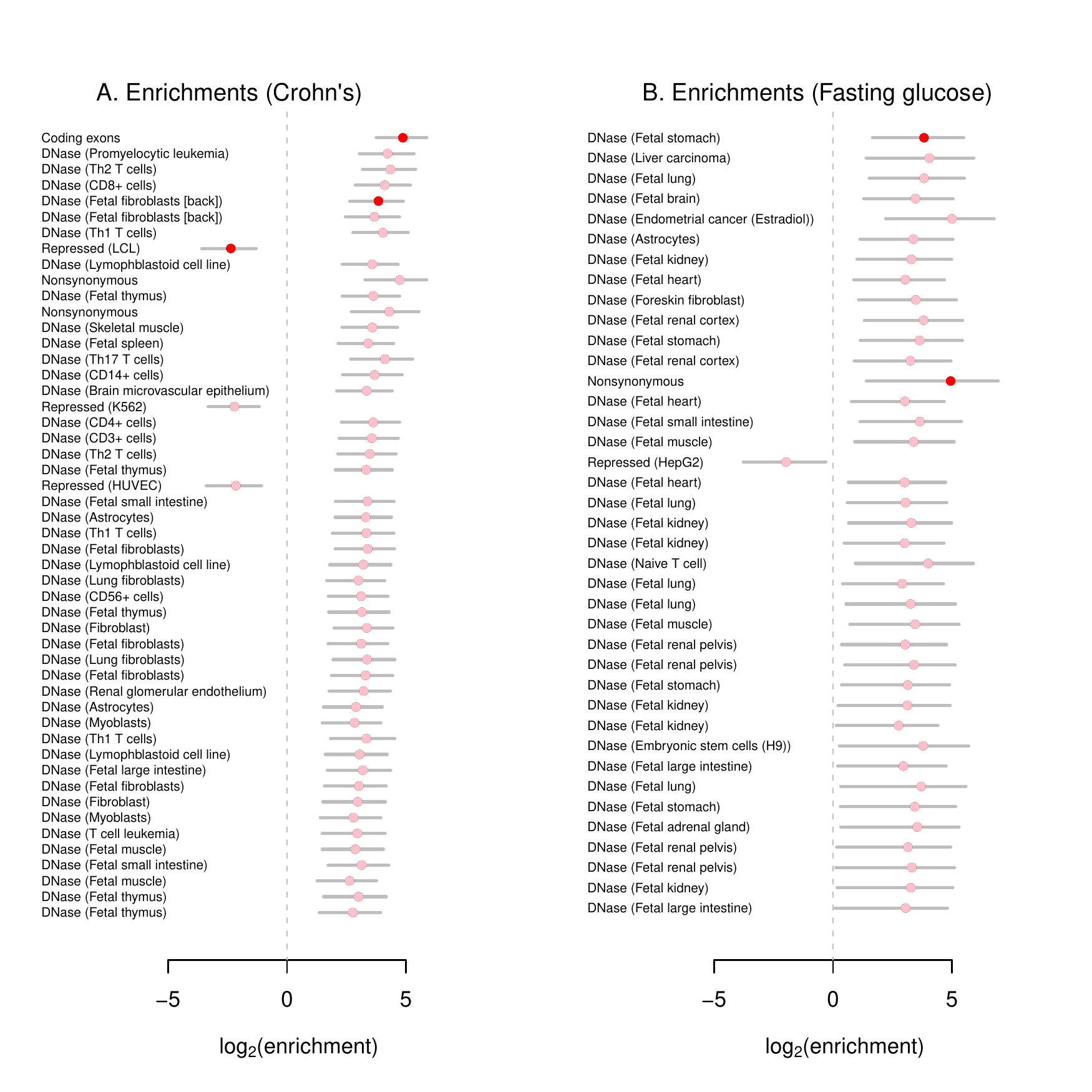}
\caption{. \textbf{Annotation effects in the Crohn's disease and fasting glucose data.} We estimated an enrichment parameter for each annotation individually in the GWAS for \textbf{A.} Crohn's disease and \textbf{B.} fasting glucose. Shown are the maximum likelihood estimates and 95\% confidence intervals. Annotations are ranked according to how much each improves the fit of the model; shown are the 50 annotations that most improve the model (or if there were less than 50 significant annotations, all of the significant annotations). In red are the annotations included in the combined model, and in pink are annotations that are statistically equivalent to those in the combined model.}\label{fig_cd+fg}
\end{center}
\end{figure}

\begin{figure}
\begin{center}
\includegraphics[scale = 0.8]{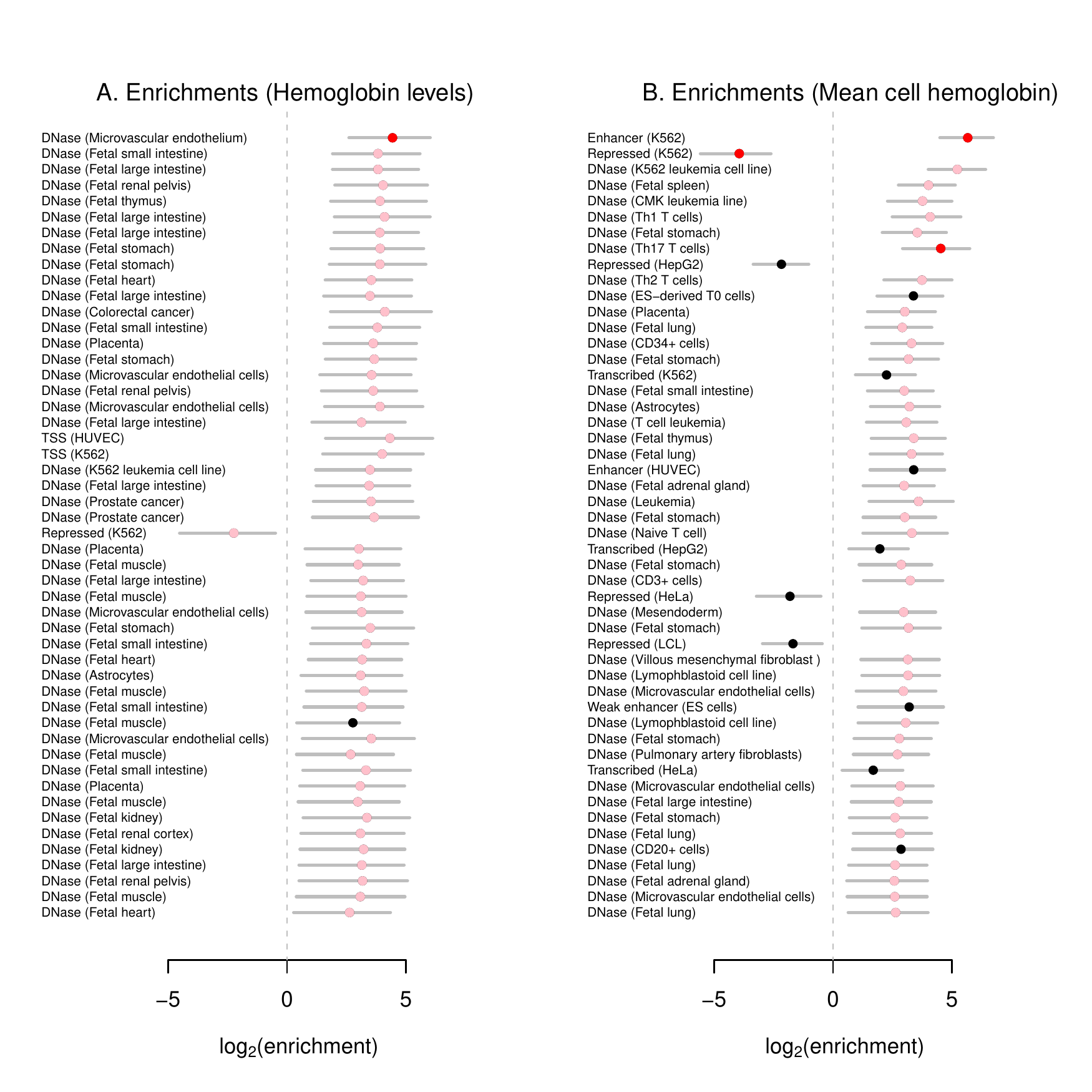}
\caption{. \textbf{Annotation effects in the red blood cell data.} We estimated an enrichment parameter for each annotation individually in the GWAS for \textbf{A.} hemoglobin levels and \textbf{B.} mean cellular hemoglobin. Shown are the maximum likelihood estimates and 95\% confidence intervals. Annotations are ranked according to how much each improves the fit of the model; shown are the 50 annotations that most improve the model (or if there were less than 50 significant annotations, all of the significant annotations). In red are the annotations included in the combined model, and in pink are annotations that are statistically equivalent to those in the combined model.}\label{fig_hb+mch}
\end{center}
\end{figure}

\begin{figure}
\begin{center}
\includegraphics[scale = 0.8]{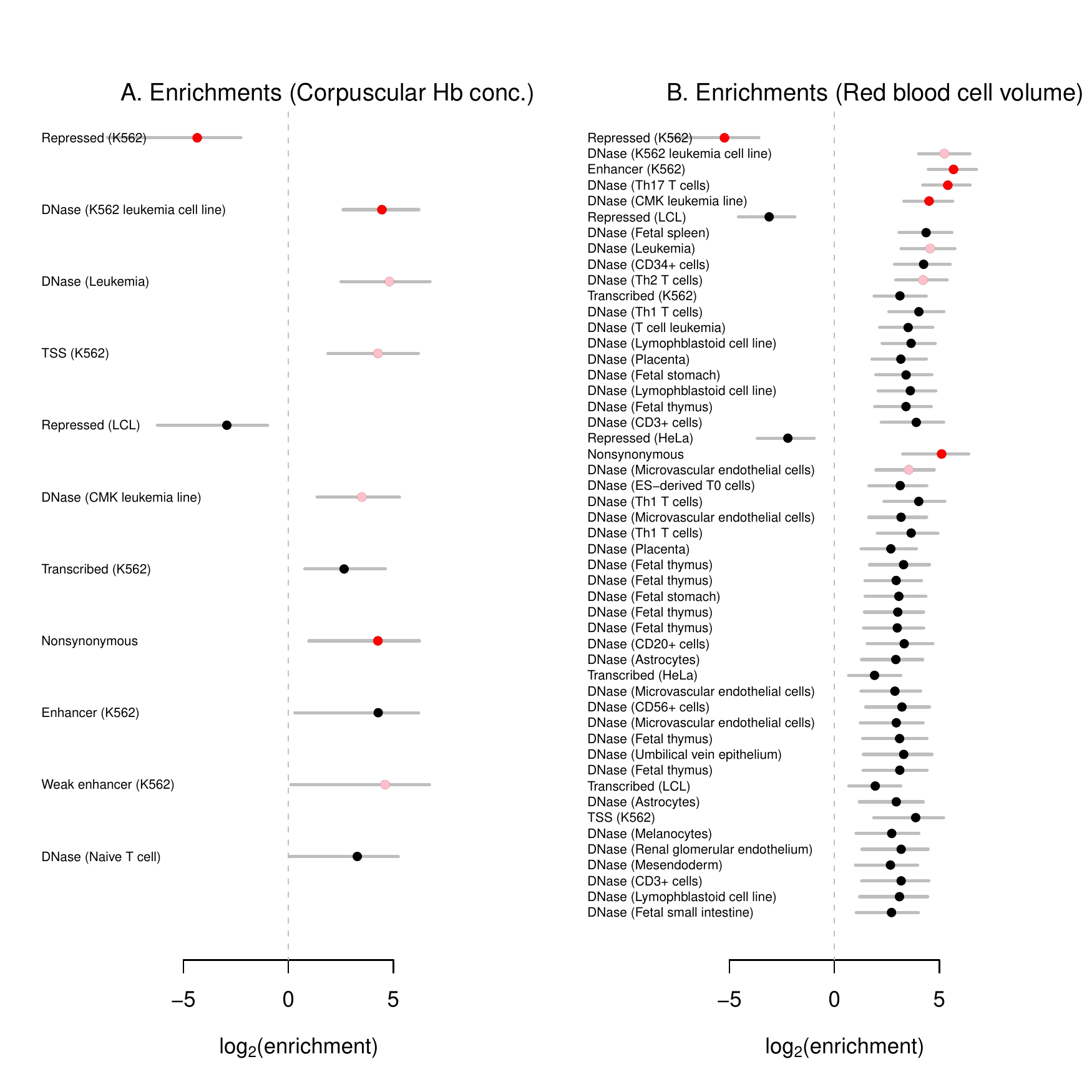}
\caption{. \textbf{Annotation effects in the red blood cell data.} We estimated an enrichment parameter for each annotation individually in the GWAS for \textbf{A.} mean corpuscular hemoglobin concentration and \textbf{B.} mean red cell volume. Shown are the maximum likelihood estimates and 95\% confidence intervals. Annotations are ranked according to how much each improves the fit of the model; shown are the 50 annotations that most improve the model (or if there were less than 50 significant annotations, all of the significant annotations). In red are the annotations included in the combined model, and in pink are annotations that are statistically equivalent to those in the combined model.}\label{fig_mchc+mcv}
\end{center}
\end{figure}

\begin{figure}
\begin{center}
\includegraphics[scale = 0.8]{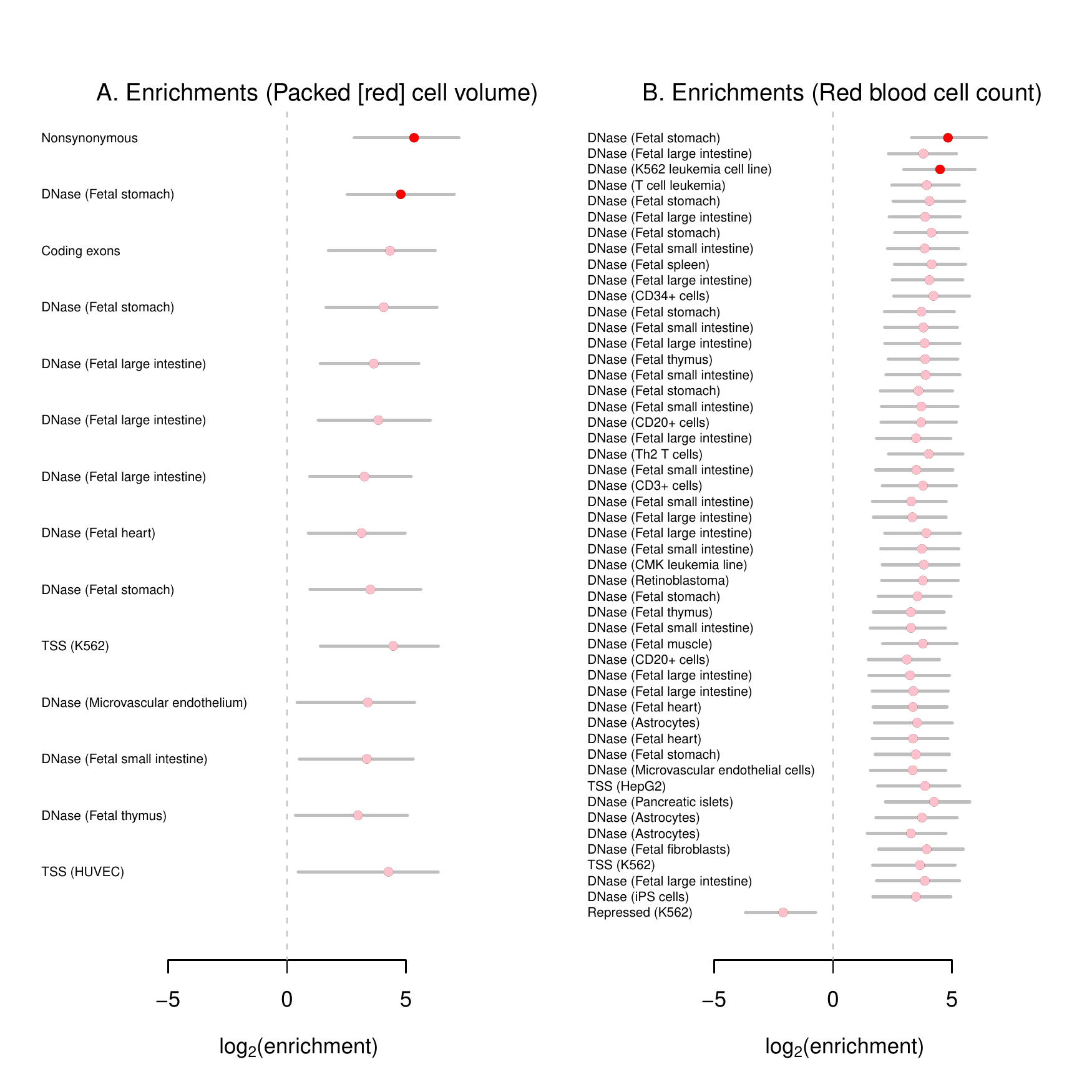}
\caption{. \textbf{Annotation effects in the red blood cell data.} We estimated an enrichment parameter for each annotation individually in the GWAS for \textbf{A.} packed cell volume and \textbf{B.} mean red cell count. Shown are the maximum likelihood estimates and 95\% confidence intervals. Annotations are ranked according to how much each improves the fit of the model; shown are the 50 annotations that most improve the model (or if there were less than 50 significant annotations, all of the significant annotations). In red are the annotations included in the combined model, and in pink are annotations that are statistically equivalent to those in the combined model.}\label{fig_pcv+rbc}
\end{center}
\end{figure}

\begin{figure}
\begin{center}
\includegraphics[scale = 0.8]{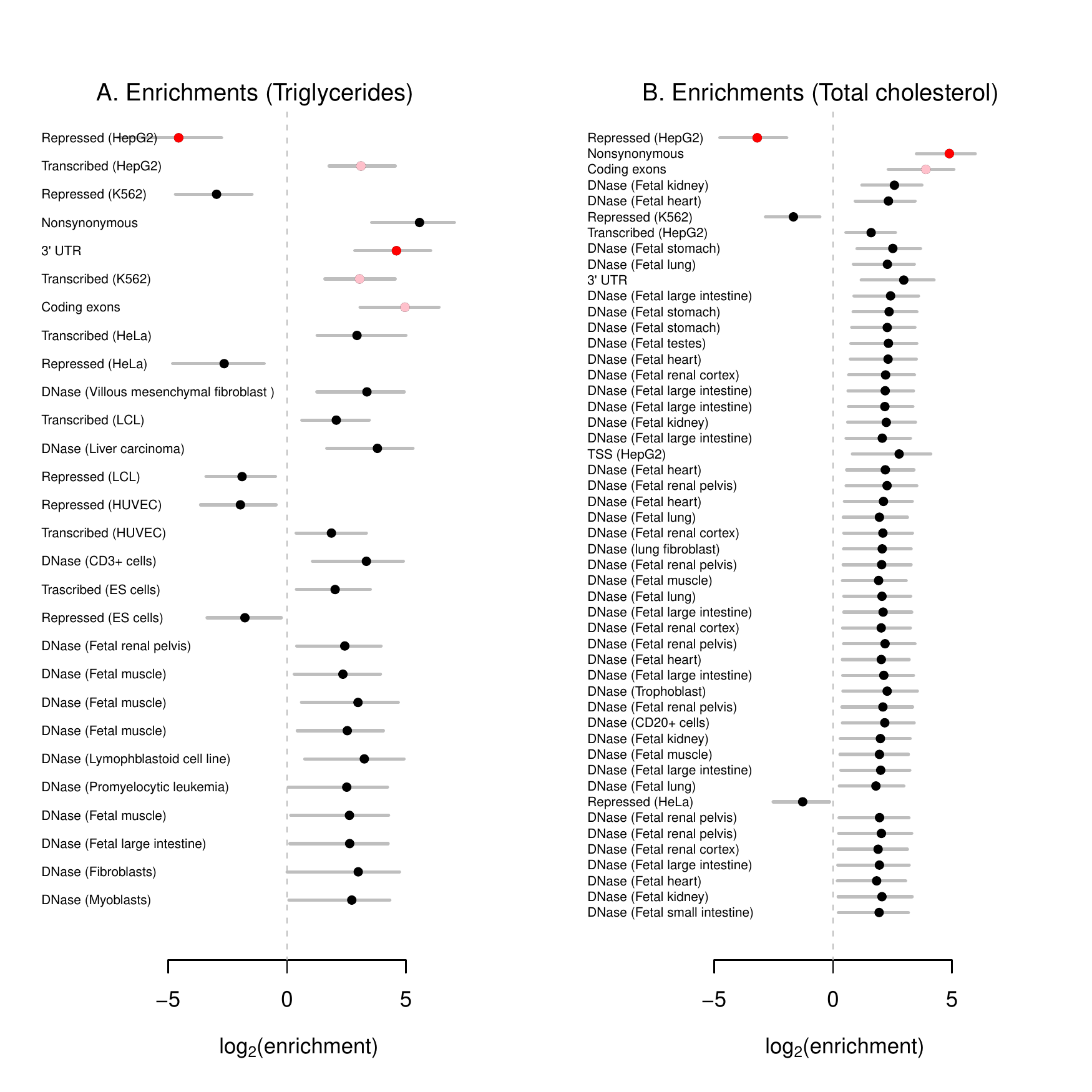}
\caption{. \textbf{Annotation effects in the lipids data.} We estimated an enrichment parameter for each annotation individually in the GWAS for \textbf{A.} triglyceride levels and \textbf{B.} total cholesterol. Shown are the maximum likelihood estimates and 95\% confidence intervals. Annotations are ranked according to how much each improves the fit of the model; shown are the 50 annotations that most improve the model (or if there were less than 50 significant annotations, all of the significant annotations). In red are the annotations included in the combined model, and in pink are annotations that are statistically equivalent to those in the combined model.}\label{fig_tg+tc}
\end{center}
\end{figure}

\begin{figure}
\begin{center}
\includegraphics[scale = 0.8]{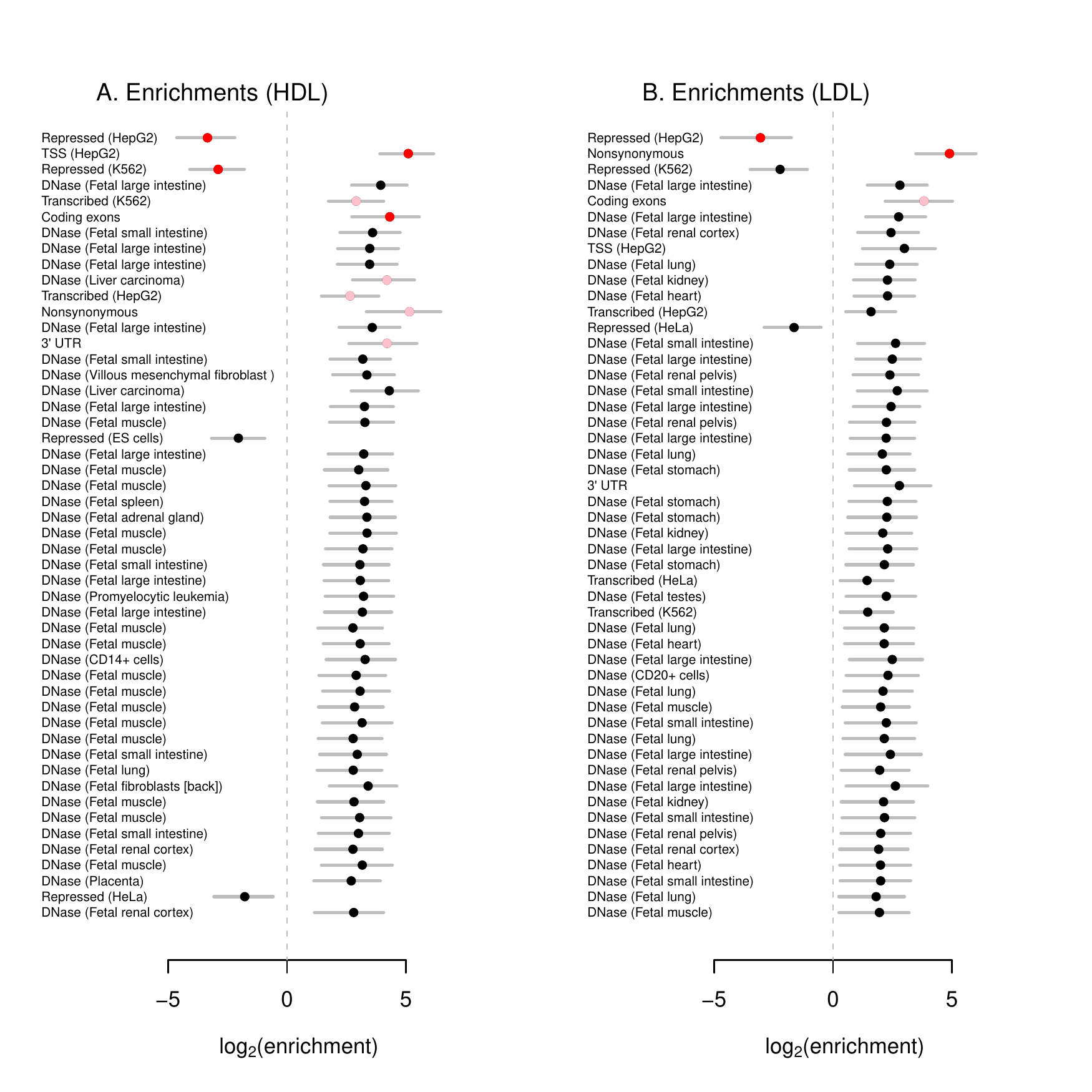}
\caption{. \textbf{Annotation effects in the lipids data.} We estimated an enrichment parameter for each annotation individually in the GWAS for \textbf{A.} HDL levels and \textbf{B.} LDL levels. Shown are the maximum likelihood estimates and 95\% confidence intervals. Annotations are ranked according to how much each improves the fit of the model; shown are the 50 annotations that most improve the model (or if there were less than 50 significant annotations, all of the significant annotations). In red are the annotations included in the combined model, and in pink are annotations that are statistically equivalent to those in the combined model.}\label{fig_ldl+hdl}
\end{center}
\end{figure}

\begin{figure}
\begin{center}
\includegraphics[scale = 0.8]{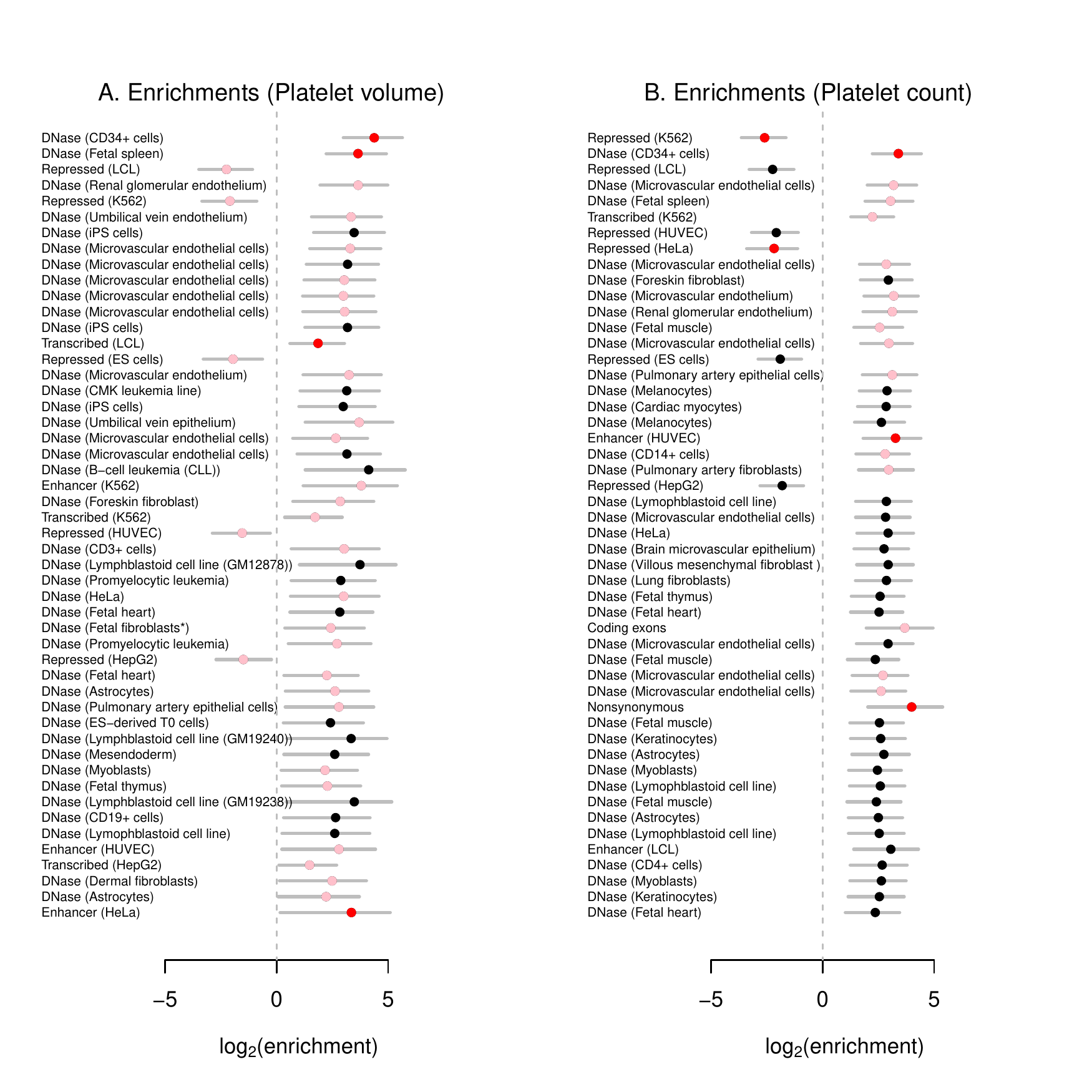}
\caption{. \textbf{Annotation effects in the platelet data.} We estimated an enrichment parameter for each annotation individually in the GWAS for \textbf{A.} mean platelet volume and \textbf{B.} platelet count. Shown are the maximum likelihood estimates and 95\% confidence intervals. Annotations are ranked according to how much each improves the fit of the model; shown are the 50 annotations that most improve the model (or if there were less than 50 significant annotations, all of the significant annotations). In red are the annotations included in the combined model, and in pink are annotations that are statistically equivalent to those in the combined model.}\label{fig_mpv+plt}
\end{center}
\end{figure}

\begin{figure}
\begin{center}
\includegraphics[scale = 0.8]{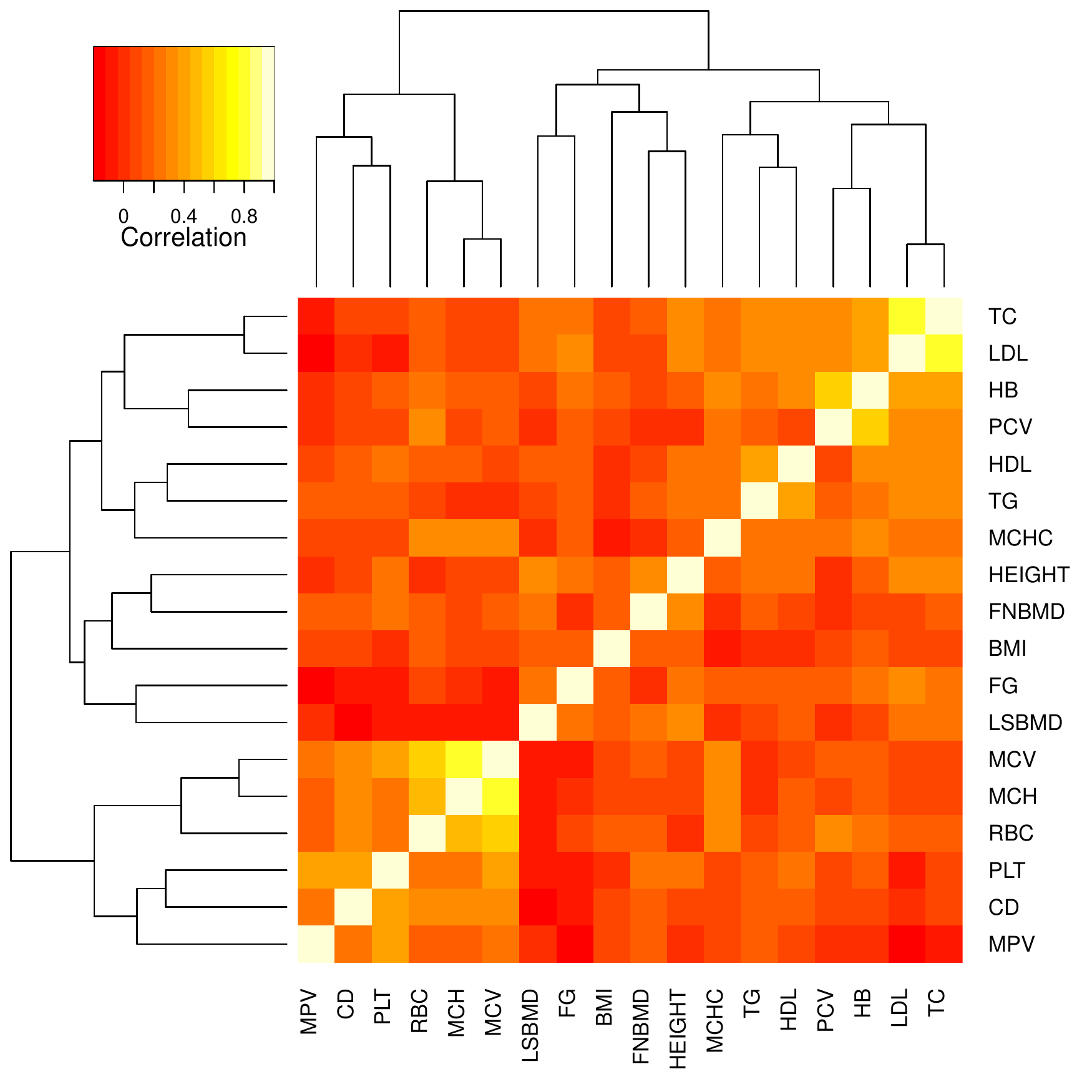}
\caption{. \textbf{Correlated patterns of enrichment across traits.} We estimated an enrichment parameter for each of 450 annotations for each of the 18 traits. For each pair of traits, we then estimated the Spearman correlation coefficient between the enrichment parameters. Plotted are these correlation coefficients. Orders of rows and columns were chosen by hierarchical clustering in R \citep{R-Core-Team:2013aa}.}\label{fig_phenocor}
\end{center}
\end{figure}

\begin{figure}
\begin{center}
\includegraphics[scale = 1]{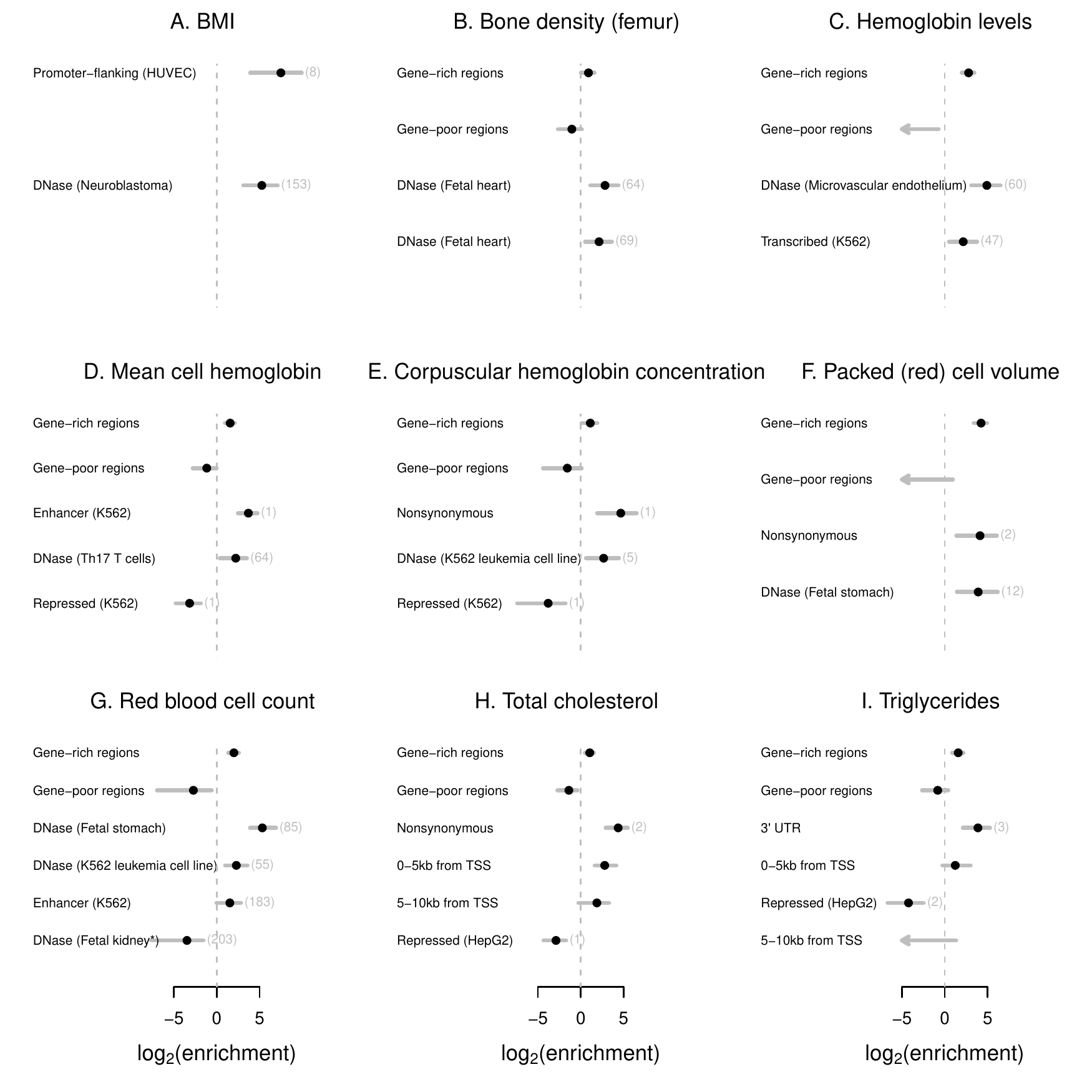}
\caption{. \textbf{Combined models for nine traits.} For each trait, we built a combined model of annotations using the algorithm presented in the Methods from the main text. Shown are the maximum likelihood estimates and 95\% confidence intervals for all annotations included in each model. Note that though these are the maximum likelihood estimates, model choice was done using a penalized likelihood. In parentheses next to each annotation (expect for those relating to distance to transcription start sites), we show the total number of annotations that are statistically equivalent to the included annotation in a conditional analysis. For the other nine traits, see Figure 4 in the main text. *This annotation of DNase-I hypersensitive sites in fetal kidney (renal pelvis) has a positive effect when treated alone; see Supplementary Text for discussion.}\label{fig_allpheno}
\end{center}
\end{figure}

\begin{figure}
\begin{center}
\includegraphics[scale = 0.8]{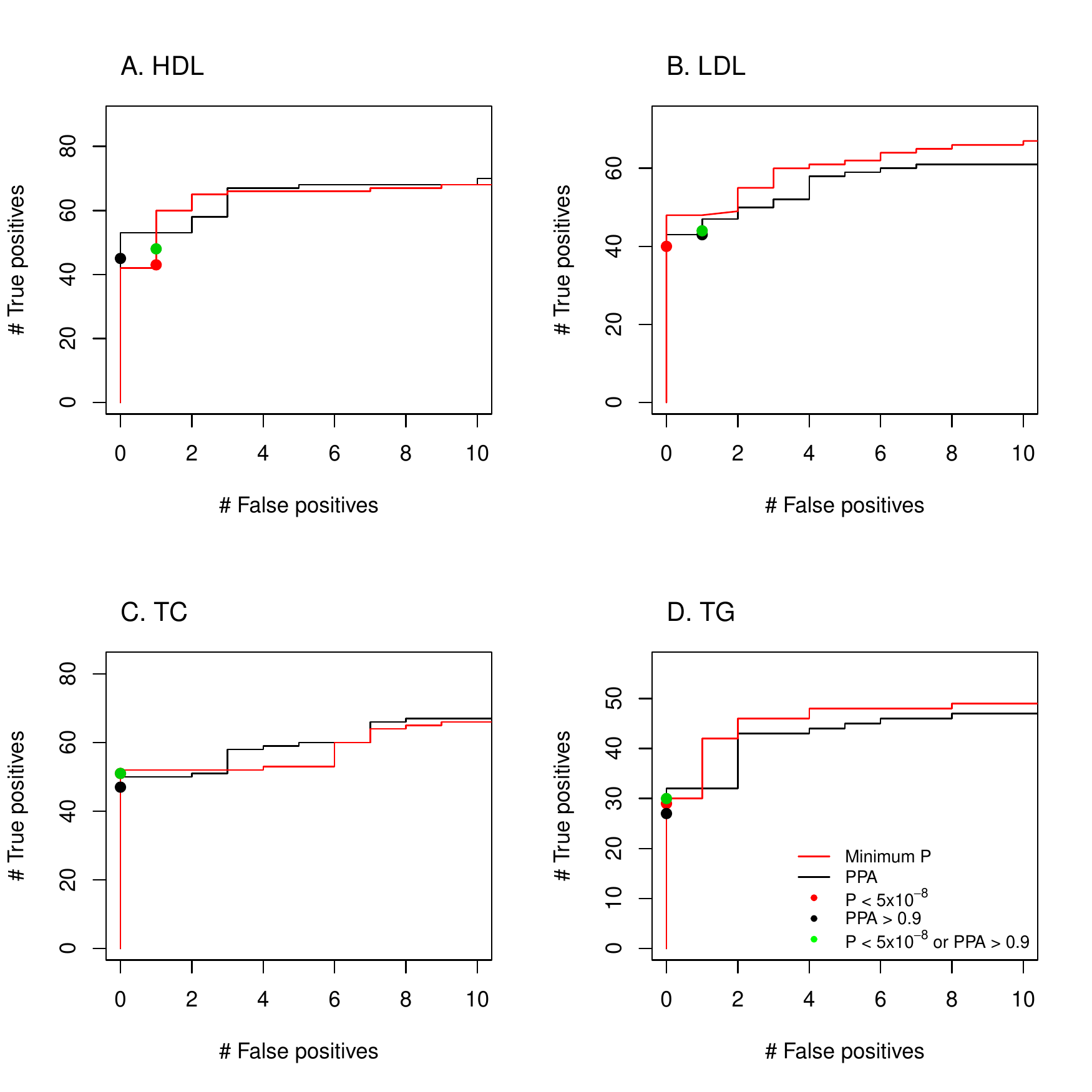}
\caption{. \textbf{Calibrating a PPA threshold similar to a P-value threshold.} For each of the four phenotypes in the lipids data, we plot the number of ``true positives" and ``false positives" obtained by different statistical thresholds; see Supplementary Text for details. Points show the positions of the thresholds used in the paper.}\label{fig_roc}
\end{center}
\end{figure}

\begin{figure}
\begin{center}
\includegraphics[scale = 0.8]{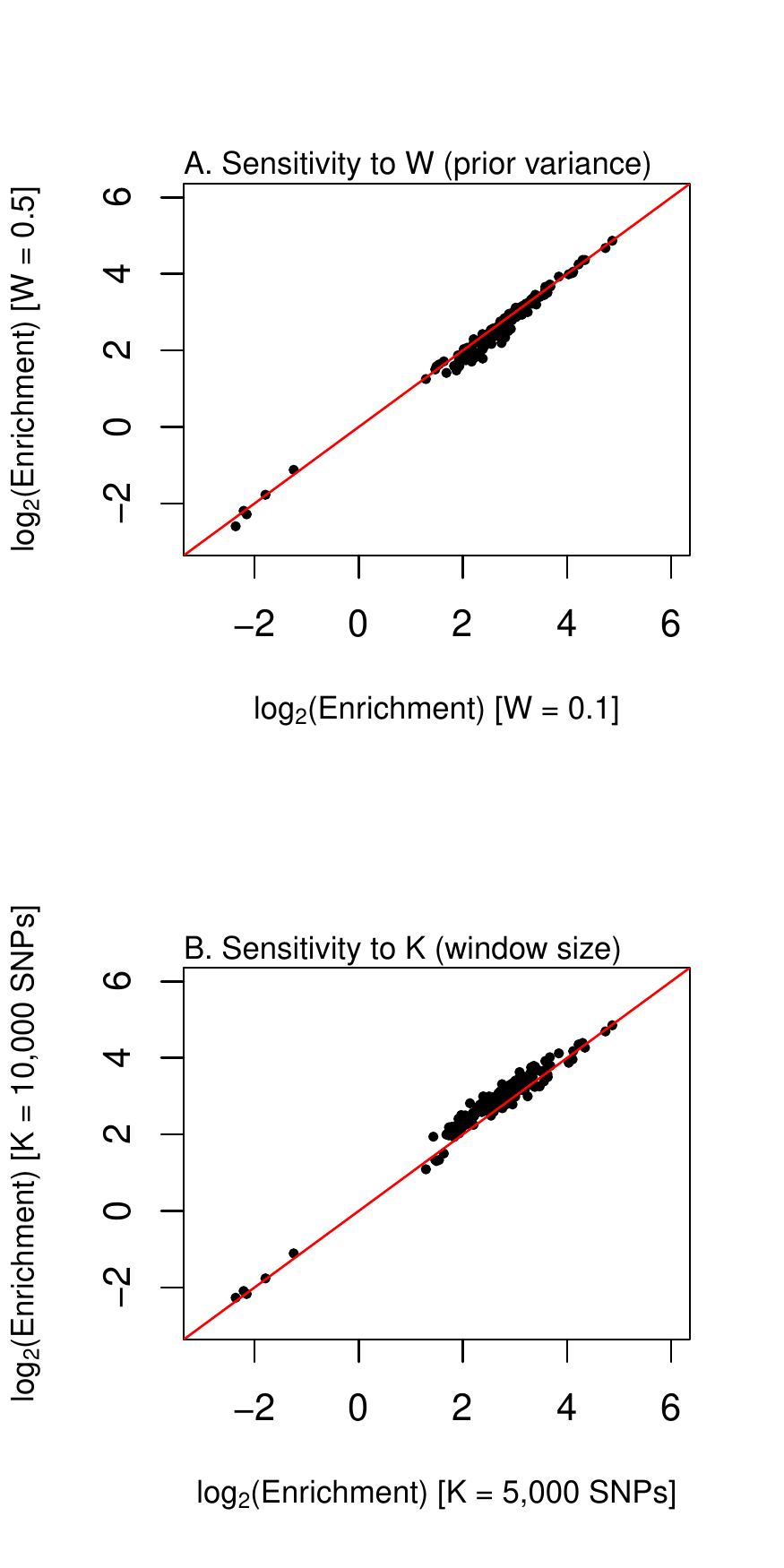}
\caption{. \textbf{Robustness of parameter estimates to preset parameters. A. Prior variance on effect size.} We estimated an enrichment parameter for each annotation in Crohn's disease using prior variances of 0.1 or 0.5. Shown are the estimates for all annotations with 95\% confidence intervals that did not overlap 0 in at least one of the two runs. In red is the $y =x$ line. \textbf{B. Window size.} We estimated an enrichment parameter for each annotation in Crohn's disease using window sizes of 5,000 and 10,000 SNPs. Shown are the estimates for all annotations with 95\% confidence intervals that did not overlap 0 in at least one of the two runs. In red is the $y =x$ line. }\label{fig_prior}
\end{center}
\end{figure}

\clearpage
\begin{table}[!tbp]
\small
\begin{center}
\begin{tabular}{ c| c | c |  }
Phenotype & $\lambda_{GC}$ (before imputation) & $\lambda_{GC}$ (after imputation) \\ \hline
Height & 1.04 & 0.99 \\ \hline
BMI & 1.04 & 0.97 \\ \hline
BMD (femoral neck) & 1.0 & 0.92 \\ \hline
BMD (lumbar spine) & 1.0 & 0.93 \\ \hline
Crohn's & 1.27 & 0.71 \\ \hline
FG & 1.08 & 0.97 \\ \hline
HB & 1.07 & 0.99  \\ \hline
MCH & 1.13 & 1.0 \\ \hline
MCHC & 1.07 & 0.85  \\  \hline
MCV & 1.13 & 1.0  \\  \hline
PCV & 1.09 & 0.97 \\ \hline
RBC & 1.14 & 1.01  \\ \hline
TC & 1.0 & 0.93 \\ \hline
TG & 1.0 & 0.92 \\ \hline
HDL & 1.0 & 0.94  \\ \hline
LDL & 1.0 & 0.93  \\ \hline
PLT & 1.08 & 1.01 \\ \hline
MPV & 1.04 & 0.96  \\ \hline
\hline 
\end{tabular}
\end{center}
\caption{: \textbf{Genomic control inflation factors before and after imputation.} We show $\lambda_{GC}$ \citep{bacanu2002association} before and after imputation for all 18 GWAS included in this study.} \label{lambda_table}
\end{table}

\begin{table}[!tbp]
\small
\begin{center}
\begin{tabular}{ c| c  }
Phenotype & Proportion [95\% CI] \\ \hline
BMI & 0.022 [0.013, 0.032] \\ \hline
FNBMD  & 0.028 [0.019, 0.040] \\ \hline
LSBMD & 0.028 [0.019, 0.041] \\ \hline
Crohn's & 0.078 [0.059, 0.10]\\ \hline
FG & 0.020 [0.012, 0.03] \\ \hline
HB & 0.010 [0.006, 0.015] \\ \hline
HDL & 0.034 [0.026, 0.044] \\ \hline
Height & 0.131 [0.111, 0.153] \\ \hline
LDL & 0.034 [0.026, 0.045] \\ \hline
MCH & 0.035 [0.025, 0.047] \\ \hline
MCHC & 0.018 [0.011, 0.027] \\ \hline
MCV & 0.046 [0.034, 0.059]\\ \hline
MPV & 0.025 [0.017, 0.035] \\ \hline
PCV & 0.003 [0.002, 0.005] \\ \hline
PLT & 0.036 [0.028, 0.047] \\ \hline
RBC & 0.023 [0.016, 0.033]  \\ \hline
TC & 0.052 [0.040, 0.067]\\ \hline
TG & 0.023 [0.015, 0.032]\\ \hline
\hline 
\end{tabular}
\end{center}
\caption{: \textbf{Estimates of the fraction of regions containing an associated SNP for each phenotype.} We show the estimates of $\frac{1}{1+e^{-\kappa}}$, the proportion of regions from the middle third of the distribution of gene density that contain associated SNPs (see Equation 7 in the main text), along with the 95\% confidence interval of this parameter.} \label{kappa_table}
\end{table}

\clearpage
\begin{table}[!tbp]
\small
\begin{center}
\begin{tabular}{ c| C{4cm}| c | c | c|  }
Annotation & Description & log$_2$(Effect) [95\% CI] & Penalized effect &  Marginal effect [95\% CI]  \\ \hline
BE2\_C-DS14625 & DNAse-I in BE(2)-C neuroblastoma cell line & 5.25 [3.09, 7.10] & 5.15 & 5.60 [3.51, 7.47]\\ \hline
HUVEC PF & Genome segmentation in HUVEC cells: promoter-flanking & 7.47 [3.90,9.91] & 7.18 & 8.51 [5.41, 10.62] \\ \hline
\hline 
\end{tabular}
\end{center}
\caption{: \textbf{Combined model learned for BMI.} Shown are the exact annotation names and parameters learned for BMI, along with the penalized effect sizes and the effect of each annotation in a single-annotation model.} \label{bmi_table}
\end{table}

\begin{table}[!tbp]
\small
\begin{center}
\begin{tabular}{ c| C{4cm}| c | c | c|  }
Annotation & Description &  log$_2$(Effect) [95\% CI] & Penalized effect &  Marginal effect [95\% CI]  \\ \hline
High gene density & Regional annotation: top 1/3 of gene density & 0.89 [0.01, 1.62] & 0.88 & NA \\ \hline
Low gene density & Regional annotation: bottom 1/3 of gene density  & -1.05 [-2.68, 0.12] & -0.95 & NA \\ \hline
fHeart-DS12810 & DNase-I in fetal heart & 2.83 [1.11, 4.40] & 2.45 & 4.83 [3.08, 6.43] \\ \hline
fHeart-DS16621 & DNase-I in fetal heart & 2.12 [0.50, 3.64] & 2.21 & 4.47 [2.76, 6.03] \\ \hline
\hline 
\end{tabular}
\end{center}
\caption{: \textbf{Combined model learned for bone mineral density (femur).} Shown are the exact annotation names and parameters learned for FNBMD, along with the penalized effect sizes and the effect of each annotation in a single-annotation model.} \label{fnbmd_table}
\end{table}

\begin{table}[!tbp]
\small
\begin{center}
\begin{tabular}{ c| C{4cm}| c | c | c|  }
Annotation & Description &  log$_2$(Effect)  [95\% CI] & Penalized effect &  Marginal effect [95\% CI]  \\ \hline
High gene density & Regional annotation: top 1/3 of gene density & 0.52 [-0.40, 1.27] & 0.53 & NA \\ \hline
Low gene density & Regional annotation: bottom 1/3 of gene density  & -1.49 [-3.65, -0.13] & -1.33 & NA \\ \hline
HSMM\_D-DS15542 & DNase-I in skeletal muscle myoblasts & 4.23 [2.24, 5.97] & 3.75 & 3.90 [1.98, 5.58] \\ \hline
\hline 
\end{tabular}
\end{center}
\caption{: \textbf{Combined model learned for bone mineral density (spine).} Shown are the exact annotation names and parameters learned for LSBMD, along with the penalized effect sizes and the effect of each annotation in a single-annotation model. } \label{lsbmd_table}
\end{table}

\begin{table}[!tbp]
\footnotesize
\begin{center}
\begin{tabular}{ c| C{3.5cm}| c | c | c|  }
Annotation & Description &  log$_2$(Effect) Effect [95\% CI] & Penalized effect &  Marginal effect [95\% CI]  \\ \hline
High gene density & Regional annotation: top 1/3 of gene density & 1.18 [0.61, 1.72] & 1.18 & NA \\ \hline
Low gene density & Regional annotation: bottom 1/3 of gene density  & -2.18 [-4.10, -0.92] & -2.03 & NA \\ \hline
fSkin\_fibro\_upper\_back-DS19696 & DNase-I in fetal skin fibroblasts from the upper back & 5.21 [4.08, 6.20] & 4.78 &  3.84 [2.63, 4.89] \\ \hline
gm12878.combined.R & Genome segmentation of GM12878: repressed & -1.83 [-3.06, -0.78]& -1.79 & -2.35 [-4.50, -1.05]\\ \hline
fSkin\_fibro\_abdomen-DS19561 & DNase-I in fetal skin fibroblasts from abdomen & -2.34 [-3.85, -1.18] & -1.86 & 2.77 [1.27, 3.94] \\ \hline
huvec.combined.T & Genome segmentation of HUVEC: transcribed & 1.20 [0.25, 2.15] & 1.17 & 1.63 [0.61, 2.65] \\ \hline
Distance to TSS [0-5 kb]& From 0-5 kb from a TSS& 1.18 [0.17, 2.15] & 1.17 & NA \\ \hline
Distance to TSS [5-10 kb]& From 5-10 kb from a TSS & 0.45 [-1.38, 1.75] & 0.40 & NA  \\ \hline
\hline 
\end{tabular}
\end{center}
\caption{: \textbf{Combined model learned for Crohn's disease.} Shown are the exact annotation names and parameters learned for Crohn's disease, along with the penalized effect sizes and the effect of each annotation in a single-annotation model.} \label{cd_table}
\end{table}

\begin{table}[!tbp]
\footnotesize
\begin{center}
\begin{tabular}{ c| C{3.5cm}| c | c | c|  }
Annotation & Description & log$_2$(Effect)  [95\% CI] & Penalized effect &  Marginal effect [95\% CI]  \\ \hline
fStomach-DS17878 & DNase-I in fetal stomach & 3.66 [1.66, 5.31] & 3.62 & 3.82 [1.66, 5.50]   \\ \hline
Nonsynonymous & nonsynonymous SNPs & 4.28 [1.53, 6.10]& 4.13 & 4.95 [1.40, 6.95] \\ \hline
Distance to TSS [0-5 kb]& From 0-5 kb from a TSS& 1.83 [0.22, 3.40] & 1.75 & NA \\ \hline
Distance to TSS [5-10 kb]& From 5-10 kb from a TSS & 2.68 [0.76, 4.28] & 2.54 & NA  \\ \hline
\hline 
\end{tabular}
\end{center}
\caption{: \textbf{Combined model learned for fasting glucose.} Shown are the exact annotation names and parameters learned for FG, along with the penalized effect sizes and the effect of each annotation in a single-annotation model.} \label{fg_table}
\end{table}

\begin{table}[!tbp]
\footnotesize
\begin{center}
\begin{tabular}{ c| C{3.5cm}| c | c | c|  }
Annotation & Description &  log$_2$(Effect) 95\% CI] & Penalized effect &  Marginal effect [95\% CI]  \\ \hline
High gene density & Regional annotation: top 1/3 of gene density & 2.78 [1.98, 3.46] & 2.80 & NA \\ \hline
Low gene density & Regional annotation: bottom 1/3 of gene density  & -40.6 [$-\inf$, -0.72] & -5.44 & NA \\ \hline
HMVEC\_dAd-DS12957 & DNase-I in microvascular endothelium & 4.91 [3.09, 6.52] & 4.86 & 4.43 [2.60, 6.02]   \\ \hline
k562.combined.T & Genome segmentation of K562: transcribed & 2.15 [0.49, 3.77]& 2.12 & 1.82 [0.01, 3.55] \\ \hline
\hline 
\end{tabular}
\end{center}
\caption{: \textbf{Combined model learned for hemoglobin levels.} Shown are the exact annotation names and parameters learned for hemoglobin levels, along with the penalized effect sizes and the effect of each annotation in a single-annotation model.} \label{hb_table}
\end{table}

\begin{table}[!tbp]
\footnotesize
\begin{center}
\begin{tabular}{ c| C{3.5cm}| c | c | c|  }
Annotation & Description & log$_2$(Effect) [95\% CI] & Penalized effect &  Marginal effect [95\% CI]  \\ \hline
High gene density & Regional annotation: top 1/3 of gene density & 1.69 [1.13, 2.19] & 1.56 & NA \\ \hline
Low gene density & Regional annotation: bottom 1/3 of gene density  & -1.17 [-0.13, 0.69] & -0.20 & NA \\ \hline
hepg2.combined.R & Genome segmentation of HepG2: repressed & -1.83 [-3.12, -0.68] & -1.79 & -3.35 [-4.63, -2.19]   \\ \hline
hepg2.combined.TSS & Genome segmentation of HepG2: TSS & 3.10 [1.79, 4.20]& 2.84 & 5.09 [3.91, 6.16] \\ \hline
ens\_coding\_exons & Ensembl: coding exons & 3.16 [1.51, 4.40] & 2.73 & 4.31 [2.73, 5.55] \\ \hline
k562.combined.R &  Genome segmentation of K562: repressed & -1.43 [-2.65, -0.30] & -1.43 &  -2.90 [-4.08, -1.79] \\ \hline
\hline 
\end{tabular}
\end{center}
\caption{: \textbf{Combined model learned for HDL levels.} Shown are the exact annotation names and parameters learned for HDL, along with the penalized effect sizes and the effect of each annotation in a single-annotation model.} \label{hdl_table}
\end{table}

\begin{table}[!tbp]
\footnotesize
\begin{center}
\begin{tabular}{ c| C{3.5cm}| c | c | c|  }
Annotation & Description & log$_2$(Effect) [95\% CI] & Penalized effect &  Marginal effect [95\% CI]  \\ \hline
High gene density & Regional annotation: top 1/3 of gene density & 1.50 [1.13, 1.86] & 1.49 & NA \\ \hline
Low gene density & Regional annotation: bottom 1/3 of gene density  & -0.95 [-1.62, -0.36] & -0.94 & NA \\ \hline
helas3.combined.R & Genome segmentation of HeLa: repressed & -1.50 [-2.39, -0.71] & -1.50 & -2.74 [-3.78, -1.85]   \\ \hline
fMuscle\_lower\_limb-DS18174 & DNase-I in fetal muscle from lower limb & 2.27 [1.50, 3.02] & 2.24 & 3.61 [2.81, 4.40] \\ \hline
Nonsynonymous & Nonsynonymous SNPs & 3.74 [2.55, 4.65] & 3.58 & 4.27 [2.77, 5.32] \\ \hline
fLung-DS15573 & DNase-I in fetal lung & 2.09 [1.30, 2.80] & 2.05 & 3.77 [2.97, 4.50] \\ \hline
huvec.combined.T & Genome segmentation of HUVEC: transcribed & 1.27 [0.52, 1.96] & 1.24 & 1.63 [0.89, 2.34] \\ \hline
ens\_utr3\_exons & Ensembl: 3' UTRs & 1.57 [0.00, 2.64] & 1.54 & 2.93 [1.34, 3.98] \\ \hline
\hline 
\end{tabular}
\end{center}
\caption{: \textbf{Combined model learned for height.} Shown are the exact annotation names and parameters learned for height, along with the penalized effect sizes and the effect of each annotation in a single-annotation model. } \label{height_table}
\end{table}

\begin{table}[!tbp]
\footnotesize
\begin{center}
\begin{tabular}{ c| C{3.5cm}| c | c | c|  }
Annotation & Description & log$_2$(Effect) [95\% CI] & Penalized effect &  Marginal effect [95\% CI]  \\ \hline
High gene density & Regional annotation: top 1/3 of gene density & 1.77 [1.21, 2.27] & 1.72 & NA \\ \hline
Low gene density & Regional annotation: bottom 1/3 of gene density  & -0.72 [-1.98, 0.25] & -0.71 & NA \\ \hline
hepg2.combined.R & Genome segmentation of HepG2: repressed & -2.78 [-4.36, -1.51] & -2.70 & -3.04 [-4.70, -1.76] \\ \hline
Nonsynonymous & Nonsynonymous SNPs & 4.24 [2.74, 5.40] & 3.97 & 4.89 [3.48, 6.02] \\ \hline
Distance to TSS [0-5 kb]& From 0-5 kb from a TSS& 3.13 [1.96, 4.56] & 2.84 & NA \\ \hline
Distance to TSS [5-10 kb]& From 5-10 kb from a TSS & 1.63 [-0.65, 3.12] & 1.17 & NA  \\ \hline
\hline 
\end{tabular}
\end{center}
\caption{: \textbf{Combined model learned for LDL levels.} Shown are the exact annotation names and parameters learned for LDL, along with the penalized effect sizes and the effect of each annotation in a single-annotation model.} \label{ldl_table}
\end{table}

\begin{table}[!tbp]
\footnotesize
\begin{center}
\begin{tabular}{ c| C{3.5cm}| c | c | c|  }
Annotation & Description & log$_2$(Effect) [95\% CI] & Penalized effect &  Marginal effect [95\% CI]  \\ \hline
High gene density & Regional annotation: top 1/3 of gene density & 1.56 [0.94, 2.11] & 1.51 & NA \\ \hline
Low gene density & Regional annotation: bottom 1/3 of gene density  & -1.17 [-2.80, -0.01] & -1.10 & NA \\ \hline
k562.combined.E & Genome segmentation of K562: enhancers & 3.68 [2.47, 4.75] & 3.53 & 5.67 [4.49, 6.74] \\ \hline
k562.combined.R & Genome segmentation of K562: repressed & -3.17 [-4.80, -1.86] & -2.97 & -3.94 [-5.57, -2.61] \\ \hline
hTH17-DS11039 & DNase-I in Th17 T cells & 2.21 [0.35, 3.51]  & 2.06 & 4.53 [2.93, 5.74] \\ \hline
\hline 
\end{tabular}
\end{center}
\caption{: \textbf{Combined model learned for mean cell hemoglobin.} Shown are the exact annotation names and parameters learned for MCH, along with the penalized effect sizes and the effect of each annotation in a single-annotation model.} \label{mch_table}
\end{table}

\begin{table}[!tbp]
\footnotesize
\begin{center}
\begin{tabular}{ c| C{3.5cm}| c | c | c|  }
Annotation & Description & log$_2$(Effect) [95\% CI] & Penalized effect &  Marginal effect [95\% CI]  \\ \hline
High gene density & Regional annotation: top 1/3 of gene density & 1.11 [0.09, 1.93] & 1.17 & NA \\ \hline
Low gene density & Regional annotation: bottom 1/3 of gene density  & -1.57 [-4.41, 0.10] & -1.36 & NA \\ \hline
k562.combined.R & Genome segmentation of K562: repressed & -3.81 [-7.43, -1.79] & -3.42 & -4.34 [-8.94, -2.27] \\ \hline
K562-DS9767 & DNase-I in K562 cells & 2.67 [0.61, 4.44] & 2.47 & 4.46 [2.60, 6.22] \\ \hline
Nonsynonymous & Nonsynonymous SNPs & 4.66 [1.90, 6.52] & 4.03 & 4.27 [0.97, 6.25] \\ \hline
\hline 
\end{tabular}
\end{center}
\caption{: \textbf{Combined model learned for mean corpuscular hemoglobin concentration.} Shown are the exact annotation names and parameters learned for MCHC, along with the penalized effect sizes and the effect of each annotation in a single-annotation model.} \label{mchc_table}
\end{table}

\begin{table}[!tbp]
\footnotesize
\begin{center}
\begin{tabular}{ c| C{3.5cm}| c | c | c|  }
Annotation & Description & log$_2$(Effect) [95\% CI] & Penalized effect &  Marginal effect [95\% CI]  \\ \hline
High gene density & Regional annotation: top 1/3 of gene density & 1.36 [0.76, 1.86] & 1.31 & NA \\ \hline
Low gene density & Regional annotation: bottom 1/3 of gene density  & -1.51 [-3.06, -0.39] & -1.46 & NA \\ \hline
k562.combined.R & Genome segmentation of K562: repressed & -3.91 [-6.25, -2.38] & -3.69 & -5.24 [-7.76, -3.59] \\ \hline
k562.combined.E & Genome segmentation of K562: enhancer & 3.10 [1.86, 4.15] & 2.96 & 5.67 [4.47, 6.77] \\ \hline
hTH17-DS11039 & DNase-I in Th17 T cells & 2.31 [0.81, 3.48] & 2.25 & 5.40 [4.21, 6.46] \\ \hline
Nonsynonymous & Nonsynonymous SNPs & 4.54 [2.34, 5.92] & 4.13 & 5.11 [3.26, 6.39] \\ \hline
CMK-DS12393 & DNase-I in CMK leukemia line & 1.28 [0.04, 2.35] & 1.34 & 4.52 [3.30, 5.64] \\ \hline
Distance to TSS [0-5 kb]& From 0-5 kb from a TSS& 0.38 [-1.59, 0.65] & -0.33 & NA \\ \hline
Distance to TSS [5-10 kb]& From 5-10 kb from a TSS & 0.89 [-0.40, 1.83] & 0.84 & NA  \\ \hline
\hline 
\end{tabular}
\end{center}
\caption{: \textbf{Combined model learned for mean red cell volume.} Shown are the exact annotation names and parameters learned for MCV, along with the penalized effect sizes and the effect of each annotation in a single-annotation model. } \label{mcv_table}
\end{table}

\begin{table}[!tbp]
\footnotesize
\begin{center}
\begin{tabular}{ c| C{3.5cm}| c | c | c|  }
Annotation & Description & log$_2$(Effect) [95\% CI] & Penalized effect &  Marginal effect [95\% CI]  \\ \hline
High gene density & Regional annotation: top 1/3 of gene density & 1.95 [1.30, 2.52] & 1.88 & NA \\ \hline
Low gene density & Regional annotation: bottom 1/3 of gene density  & -2.06 [-4.73, -0.40] & -1.63 & NA \\ \hline
CD34-DS12274 & DNase-I in CD34+ cells & 3.02 [1.69, 4.26] & 2.76 & 4.37 [2.99, 5.64] \\ \hline
gm12878.combined.T & Genome segmentation of GM12878: transcribed & 2.35 [1.07, 3.53] & 1.83 & 1.86 [0.59, 3.04] \\ \hline
helas3.combined.E & Genome segmentation of HeLa: enhancer & 2.80 [0.75, 4.23] & 2.27 & 3.35 [0.16, 5.09] \\ \hline
fSpleen-DS17448 & DNase-I in fetal spleen & 1.93 [0.59, 3.15] & 1.88 & 3.65 [2.22, 4.92] \\ \hline
\hline 
\end{tabular}
\end{center}
\caption{: \textbf{Combined model learned for mean platelet volume.} Shown are the exact annotation names and parameters learned for MPV, along with the penalized effect sizes and the effect of each annotation in a single-annotation model.} \label{mpv_table}
\end{table}

\begin{table}[!tbp]
\footnotesize
\begin{center}
\begin{tabular}{ c| C{3.5cm}| c | c | c|  }
Annotation & Description & log$_2$(Effect) [95\% CI] & Penalized effect &  Marginal effect [95\% CI]  \\ \hline
High gene density & Regional annotation: top 1/3 of gene density & 4.24 [3.36, 4.95] & 3.72 & NA \\ \hline
Low gene density & Regional annotation: bottom 1/3 of gene density  & -40.60 [-$\inf$, 0.94] & -1.92 & NA \\ \hline
Nonsynonymous & Nonsynonymous SNPs & 4.11 [1.34, 6.07]  & 3.61 & 5.34 [2.83, 7.23] \\ \hline
fStomach-DS17172 & DNase-I in fetal stomach & 3.90 [1.40, 6.17] & 3.48 & 4.78 [2.54, 7.03] \\ \hline
\hline 
\end{tabular}
\end{center}
\caption{: \textbf{Combined model learned for packed red cell volume.} Shown are the exact annotation names and parameters learned for PCV, along with the penalized effect sizes and the effect of each annotation in a single-annotation model.} \label{pcv_table}
\end{table}

\begin{table}[!tbp]
\footnotesize
\begin{center}
\begin{tabular}{ c| C{3.5cm}| c | c | c|  }
Annotation & Description & log$_2$(Effect) [95\% CI] & Penalized effect &  Marginal effect [95\% CI]  \\ \hline
High gene density & Regional annotation: top 1/3 of gene density & 1.67 [2.81, 2.64] & 2.14 & NA \\ \hline
Low gene density & Regional annotation: bottom 1/3 of gene density  & -1.63 [-3.40, -0.38] & -1.51 & NA \\ \hline
k562.combined.R & Genome segmentation in K562: repressed & -1.60 [-2.63, -0.66] & -1.60 & -2.60 [-3.65, -1.64] \\ \hline
CD34-DS12274 & DNase-I in CD34+ cells & 1.82 [0.59, 2.86] & 1.80 & 3.39 [2.24, 4.43] \\ \hline
Nonsynonymous & Nonsynonymous SNPs & 3.38 [1.31, 4.79] & 3.00 & 3.98 [2.02, 5.38] \\ \hline
huvec.combined.E& Genome segmentation in HUVEC: enhancers & 1.67 [0.16, 2.84] & 1.59 & 3.27 [1.82, 4.41] \\ \hline
helas3.combined.R &Genome segmentation in HeLa: repressed & -1.17 [-2.37, -0.13] & -1.14 & -2.18 [-3.40, -1.11] \\ \hline
\hline 
\end{tabular}
\end{center}
\caption{: \textbf{Combined model learned for platelet count.} Shown are the exact annotation names and parameters learned for PLT, along with the penalized effect sizes and the effect of each annotation in a single-annotation model.} \label{plt_table}
\end{table}

\begin{table}[!tbp]
\footnotesize
\begin{center}
\begin{tabular}{ c| C{3.5cm}| c | c | c|  }
Annotation & Description & log$_2$(Effect) [95\% CI] & Penalized effect &  Marginal effect [95\% CI]  \\ \hline
High gene density & Regional annotation: top 1/3 of gene density & 1.99 [1.33, 2.58] & 1.96 & NA \\ \hline
Low gene density & Regional annotation: bottom 1/3 of gene density  & -2.74 [-6.97, -0.59] & -2.18 & NA \\ \hline
fStomach-DS17878 & DNAse-I in fetal stomach & 5.31 [3.87, 6.91] & 4.83 & 4.83 [3.30, 6.45] \\ \hline
k562.combined.E & Genome segmentation of K562: enhancer & 1.53 [-0.04, 2.83] & 1.56 & 4.28 [1.41, 5.90] \\ \hline
fKidney\_renal\_pelvis\_R-DS18663 & DNase-I in fetal renal pelvis & -3.49 [-7.68, -1.56] & -2.80 & 2.48 [0.04, 4.17] \\ \hline
K562-DS9767 & DNase-I in K562 leukemia line & 2.28 [0.97, 3.58] & 2.25 & 4.50 [2.97, 5.97] \\ \hline
\hline 
\end{tabular}
\end{center}
\caption{: \textbf{Combined model learned for red blood cell count.} Shown are the exact annotation names and parameters learned for RBC, along with the penalized effect sizes and the effect of each annotation in a single-annotation model.} \label{rbc_table}
\end{table}

\begin{table}[!tbp]
\footnotesize
\begin{center}
\begin{tabular}{ c| C{3.5cm}| c | c | c|  }
Annotation & Description &  log$_2$(Effect) [95\% CI] & Penalized effect &  Marginal effect [95\% CI]  \\ \hline
High gene density & Regional annotation: top 1/3 of gene density & 1.05 [0.48, 1.56] & 1.04 & NA \\ \hline
Low gene density & Regional annotation: bottom 1/3 of gene density  & -1.40 [-2.74, -0.39] & -1.34 & NA \\ \hline
hepg2.combined.R & Genome segmentation of HepG2: repressed & -2.90 [-4.36, -1.72] & -2.84 & -3.19 [-4.76, -1.95] \\ \hline
Nonsynonymous & Nonsynonymous SNPs & 4.36 [2.90, 5.48] & 4.18 & 4.89 [3.51, 5.99] \\ \hline
Distance to TSS [0-5 kb]& From 0-5 kb from a TSS& 2.76 [1.62, 4.15] & 2.58 & NA \\ \hline
Distance to TSS [5-10 kb]& From 5-10 kb from a TSS & 1.88 [-0.27, 3.29] & 1.56 & NA  \\ \hline
\hline 
\end{tabular}
\end{center}
\caption{: \textbf{Combined model learned for total cholesterol.} Shown are the exact annotation names and parameters learned for total cholesterol, along with the penalized effect sizes and the effect of each annotation in a single-annotation model.} \label{tc_table}
\end{table}

\begin{table}[!tbp]
\footnotesize
\begin{center}
\begin{tabular}{ c| C{3.5cm}| c | c | c|  }
Annotation & Description & log$_2$(Effect) [95\% CI] & Penalized effect &  Marginal effect [95\% CI]  \\ \hline
High gene density & Regional annotation: top 1/3 of gene density & 1.56 [0.85, 2.18] & 1.49 & NA \\ \hline
Low gene density & Regional annotation: bottom 1/3 of gene density  & -0.82 [-2.65, 0.40] & -0.78 & NA \\ \hline
hepg2.combined.R & Genome segementation of HepG2: repressed & -4.24 [-6.68, -2.47] & -3.75 & -4.56 [-7.11, -2.76] \\ \hline
ens\_utr3\_exons & Ensembl: 3' UTRs & 3.87 [2.11, 5.28]  & 3.46 & 4.60 [2.86, 6.03] \\ \hline
\hline 
\end{tabular}
\end{center}
\caption{: \textbf{Combined model learned for triglyceride levels.} Shown are the exact annotation names and parameters learned for triglycerides, along with the penalized effect sizes and the effect of each annotation in a single-annotation model.} \label{tg_table}
\end{table}
\clearpage

\begin{table}[!tbp]
\footnotesize
\begin{center}
\begin{tabular}{ c| c | c | c | c| c|c|  }
 & \multicolumn{2}{ |c|}{PPA} &\multicolumn{2}{ |c|}{P-value} & \multicolumn{2}{ |c|}{combined} \\ \hline
Phenotype & True positives  & False positives & True positives  & False positives & True positives  & False positives \\ \hline
HDL & 45 & 0 & 43 & 1& 48 & 1 \\ \hline
LDL  & 43 & 1 & 40 & 0 & 44 & 1 \\ \hline
TC & 47 & 0 & 51 & 0 & 51 & 0  \\ \hline
TG & 27 & 0 & 29 & 0 & 30 & 0 \\ \hline
\hline 
\end{tabular}
\end{center}
\caption{: \textbf{Comparison of loci identified in the lipids data with different methods.} We ranked genomic regions in GWAS of four lipid traits according to their minimum P-value or posterior probability of association from \citet{teslovich2010biological}. We then evaluated false positives and false negatives by comparison to a larger GWAS \citep{Global-Lipids-Genetics-Consortium:2013uq}. See Supplementary Text for details.} \label{roc_table}
\end{table}

\clearpage
\begin{sidewaystable}[!tbp]
\tiny
\begin{center}
\begin{tabular}{ c| c| c | c | c|  c| c|}
trait & region (hg19) & Regional PPA & lead SNP (P-value) & Nearest gene & Successful replication (SNP, $r^2$ with lead)  \\ \hline
BMI & chr13:27,75,5426-29,745,954 & 0.94 & rs9512699 ($6\times 10^{-8}$) & MTIF3 & \citep{speliotes2010association} (rs4771122, 0.73) \\\hline
BMD (femur) & chr1:170,892,281-173,086,517 & 0.93 & rs6701929 ($2\times 10^{-7}$)& DNM3 & \citep{estrada2012genome}  (rs479336, 0.93) \\\hline
HDL & chr1:25,427,217-29,426,896 & 0.96 & rs6659176 ($1.5\times 10^{-6}$) & NR0B2 & \citep{Global-Lipids-Genetics-Consortium:2013uq} (rs12748152, 0.85)\\  \hline
HDL & chr1:93,534,311-95,828,501 & 0.93 & rs2297707 ($1\times 10^{-6}$) & TMED5 &  \citep{Global-Lipids-Genetics-Consortium:2013uq} (rs12133576, 0.79) \\\hline
HDL & chr1:108,743,042-111,481,349 & 0.97 & rs12740374 ($6 \times 10^{-8}$) & CELSR2 &  \citep{Global-Lipids-Genetics-Consortium:2013uq} (rs12740374)\\\hline
HDL & chr2:85,349,339-88,736,950 & 0.98 & rs1044973 ($1.5\times 10^{-7}$) & TGOLN2 & No (sample size not increased in \citet{Global-Lipids-Genetics-Consortium:2013uq}) \\\hline
HDL & chr10:45,535,916-50,321,467 & 0.93 & rs10900223 ($1.4 \times 10^{-7}$) & MARCH8 & \citep{Global-Lipids-Genetics-Consortium:2013uq} (rs970548, 0.99)\\\hline
MCV & chr3:139,060,509-141,377,851 &  0.98 & rs13059128 ($3.8 \times 10^{-7}$) &ZBTB38 & \citep{van2012seventy} (rs6776003, 0.48)\\\hline
MCV & chr9:134,164,493-136,620,584 & 0.90 & rs8176662 ($7.5\times 10^{-7}$) & ABO & NA \\\hline
MCV & chr20:24,615,239-30,836,608 & 0.98 & rs6088962  ($7.5\times 10^{-7}$) & BCL2L1 & NA\\\hline
TG & chr16:31,050,033-49,644,030 & 0.95 & rs1549293 ($2.7 \times 10^{-7}$) & KAT8 & \citep{Global-Lipids-Genetics-Consortium:2013uq} (rs749671, 0.80)\\\hline
LDL & chr1:91,146,258-93,672,688 & 0.97 & rs7542747 ($2.3\times 10^{-7}$) & RPAP2 & \citep{Global-Lipids-Genetics-Consortium:2013uq}  (rs4970712, 0.75) \\\hline
LDL & chr1:146,751,272-152,014,485 & 0.98 & rs267733 ($7\times 10^{-8}$) & ANXA9 &  \citep{Global-Lipids-Genetics-Consortium:2013uq} (rs267733)\\\hline
LDL &chr2:116,901,934-119,001,466 & 0.98 & rs1052639 ($6.6 \times 10^{-8}$) & DDX18 & \citep{Global-Lipids-Genetics-Consortium:2013uq} (rs10490626, 0.53) \\\hline
LDL &chr13:31,693,235-34,119,073 & 0.93 & rs4942505  ($9.8 \times 10^{-8}$) & BRCA2 & \citep{Global-Lipids-Genetics-Consortium:2013uq} (rs4942505) \\  \hline
LDL & chr17:7,456,344-9,908,665 & 0.92 & rs4791641 ($2.6 \times 10^{-7}$) & PFAS & No ($P = 1.3 \times 10^{-7}$ in  \citep{Global-Lipids-Genetics-Consortium:2013uq})\\\hline
MCHC & chr7:76,062,644-78,334,941 & 0.93 & rs58176556 ($5.4 \times 10^{-8}$) & PHTF2 & NA  \\\hline
Height & chr2:240,701,166-243,060,642 & 0.98 & rs13006939 ($3.9 \times 10^{-7}$) & SEPT2 & \citep{allen2010hundreds} (rs12694997, 0.99) \\\hline
Height & chr3:11,167,568-13,294,698 & 0.98 & rs2276749 ($3.0 \times 10^{-6}$) & VGLL2 & NA \\\hline
Height & chr3:13,294,698-15,353,840 & 0.93 & rs2597513 ($1.1 \times 10^{-7}$) & HDAC11 & \citep{allen2010hundreds} (rs2597513)\\\hline
Height & chr3:55,068,506-57,000,141 & 0.94 &rs7637449  ($1.3\times 10^{-6}$) & CCDC66 & \citep{allen2010hundreds}  (rs9835332, 0.87) \\\hline
Height & chr4:72,048-2,570,837 & 0.98 & rs3958122 ($6.0 \times 10^{-8}$) & SLBP &\citep{allen2010hundreds}  (rs2247341, 0.99) \\\hline
Height & chr5:71,376,237-73,712,303 &0.98 & rs34651 ($2.5 \times 10^{-7}$) & TNPO1 & NA \\\hline
Height & chr6:108,017,102-110,694,347 &  0.95 & rs1476387 ($2.2 \times 10^{-6}$) & SMPD2 & \citep{allen2010hundreds} (rs1046943, 0.93) \\\hline
Height & chr7:22,074,248-23,998,552 & 0.99 & rs12534093 ($5.6 \times 10^{-8}$) & IGF2BP3 &\citep{allen2010hundreds} (rs12534093)\\\hline
Height & chr7:46,327,426-48,083,339 & 0.97 & rs12538905 ($2.6 \times 10^{-7}$) & IGFBP3 & NA\\\hline
Height & chr9:87,279,007-89,667,667 & 0.90 & rs405761 ($1.3 \times 10^{-7}$) & ZCCHC6 & \citep{allen2010hundreds} (rs8181166, 0.82) \\\hline
Height & chr11:12,559,691-14,685,886 & 1.0 & rs7926971 ($7.3 \times 10^{-8}$) & TEAD1 & \citep{allen2010hundreds} (rs7926971) \\\hline
Height & chr11:14,685,886-17,491,336 & 0.93 & rs757081 ($2.2 \times 10^{-6}$) & NUCB2 & \citep{allen2010hundreds} (rs1330, 0.60) \\\hline
Height & chr15:62,349,517-64,370,301 & 0.97 & rs7178424 ($2.2 \times 10^{-7}$) & C2CD4A &\citep{allen2010hundreds} (rs7178424) \\\hline
Height & chr17:19,924,256-26,838,292 & 0.96 & rs9895199 ($3.6 \times 10^{-7}$) & KCNJ12 & \citep{allen2010hundreds}  (rs4640244, 0.79) \\\hline
Height & chr17:45,331,502-47,944,460 & 0.99 & rs9904645 ($2.2 \times 10^{-7}$)& ATP5G1 &  NA \\\hline
Height & chr22:32,075,899-33,846,972 & 0.97 & rs1012366 ($6.9 \times 10^{-8}$)& SYN3 & \citep{allen2010hundreds} (rs4821083 [not in 1000 Genomes])\\\hline
Crohn's &chr2:42,522,756-44,575,426 & 1.0 & rs17031095 ($2.6 \times 10^{-7}$)& THADA & \citep{jostins2012host} (rs10495903, 0.95) \\\hline
Crohn's & chr10:59,615,595-61,881,674 & 1.0 & rs1832556 ($2.0 \times 10^{-7}$)& IPMK &\citep{jostins2012host}  (rs2790216, 0.94) \\\hline
Crohn's & chr11:61,269,649-64,734,682 & 0.98 & rs174568 ($2.8 \times 10^{-7}$) & FADS2 &\citep{jostins2012host} (rs4246215, 0.86)\\\hline
Crohn's & chr13:99,900,420-102,096,823 & 0.94 & rs3742130 ($2.3 \times 10^{-5}$)& GPR18 & \citep{jostins2012host} (rs9557195, 0.91) \\\hline
Crohn's & chr15:67,140,517-70,199,927 & 0.93 & rs11639295 ($6.4 \times 10^{-7}$)& SMAD3 & \citep{jostins2012host} (rs17293632, 0.10) \\\hline
Crohn's & chr17:17,986,955-26,038,545 & 0.92 & rs2945406 ($4.1 \times 10^{-7}$)& KSR1 &  \citep{jostins2012host}  (rs2945412, 0.13) \\\hline
PLT & chr1:44,022,121-47,087,366 & 0.99 & rs4468203 ($3.2 \times 10^{-7}$)& GPBP1L1 & NA\\\hline 
PLT & chr9:90,221,450-92,241,847 & 0.90 & rs9410382 ($1.9 \times 10^{-6}$)& S1PR3 & NA\\\hline
PLT & chr11:32,343,164-34,501,064 & 0.93 & rs7481878 ($7.2 \times 10^{-8}$)& QSER1 & NA\\\hline
MCH & chr4:86,147,717-88,340,969 & 0.98 & rs6819155 ($2.3 \times 10^{-7}$)& APP1 & NA\\\hline
MCH & chr14:102,971,016-107,289,436 & 0.93 & rs17616316 ($1.5 \times 10^{-7}$& EIF5 & \citep{van2012seventy} (rs17616316)\\\hline
HB & chr15:75,349,145-78,654,148 & 0.90 & rs1874953 ($4.2 \times 10^{-7}$)& NRG4 & \citep{van2012seventy} (rs11072566, 0.93)\\\hline
BMD (spine) & chr17:43,556,652-46,084,026 & 0.99 & rs117504376 ($3.1 \times 10^{-7}$)& MAPT (chr17 inversion)& \citep{estrada2012genome} (rs1864325, 0.99)\\\hline
RBC & chr20:54,899,828-57,013,873 & 0.96 & rs737092 ($4.5 \times 10^{-7}$) & MIR5095 & \citep{van2012seventy} (rs737092)\\\hline
MPV & chr14:67,315,438-69,802,709 & 0.91 & rs117823369 ($3.9\times 10^{-6}$& DCAF5 & NA\\\hline
FG & chr9:111,051,626 - 112,662,634 & 0.96 & rs76817627 ($3.4\times 10^{-7}$) & FAM206A  & NA \\ \hline
\hline 
\end{tabular}
\end{center}
\caption{: \textbf{Sub-threshold associations with high posterior probability.} In each GWAS, we identified regions of the genome with a posterior probability of association greater than 0.9 but with no P-values less than $5\times10^{-8}$. Shown are the positions of these regions for each trait. See Supplementary Text for details. LD between lead SNPs and replication SNPs was computed from the 1000 Genomes Project haplotypes in Europeans; the exact file versions are listed in Section \ref{imputation_sec}.} \label{assoc_table}
\end{sidewaystable}

\clearpage
\bibliography{../../Bib/bib}